\documentclass[letterpaper,titlepage,11pt]{article}
\usepackage{hyperref}

\usepackage{amssymb,amsmath,amsfonts}
\usepackage{epsfig}

\def\clock{{\count0=\time
           \divide\count0 60
           \ifnum\count0<10 0\fi\the\count0
           \multiply\count0 -60 \advance\count0 \time
           :\ifnum\count0<10 0\fi \the\count0
         }}
\newcommand{\timestamp}{{\small\vbox{\hbox{\tt\jobname.tex}
\hbox{\the\day/\the\month/\the\year, \clock}}}}

\newcommand{\CO}{\mathcal{O}}

\newcommand{\CM}{\mathcal{M}}

\newcommand{\R}{\mathbb{R}}

\newcommand{\grad}{\vec{\nabla}}
\newcommand{\el}{\kappa}

\newcommand{\nn}{\nonumber}

\newcommand{\spa}{\ , \ \ }

\newcommand{\ds}{\displaystyle}

\DeclareMathOperator{\diag}{diag}

\newtheorem{definition}{Definition}[section]

\newtheorem{theorem}[definition]{Theorem}
\newtheorem{lemma}[definition]{Lemma}
\newtheorem{corollary}[definition]{Corollary}

\newcommand{\proof}{\noindent {\bf Proof:}\ }
\newcommand{\squ}{\noindent $\square$}

\numberwithin{equation}{section}
\begin{document}

\begin{titlepage}

\rightline{\vbox{\small\hbox{\tt hep-th/0408141} }} \vskip 2.5cm

\centerline{\Large \bf Stationary and Axisymmetric Solutions}
\vskip 0.15cm
\centerline{\Large \bf of Higher-Dimensional General Relativity}

\vskip 1.6cm
\centerline{\bf Troels Harmark}
\vskip 0.5cm
\centerline{\sl The Niels Bohr Institute}
\centerline{\sl Blegdamsvej 17, 2100 Copenhagen \O, Denmark}

\vskip 0.5cm

\centerline{\small\tt harmark@nbi.dk}

\vskip 1.6cm

\centerline{\bf Abstract} \vskip 0.2cm \noindent
We study stationary and axisymmetric solutions of General Relativity,
i.e. pure gravity, in four or higher dimensions.
$D$-dimensional stationary and axisymmetric solutions are defined as having
$D-2$ commuting Killing vector fields.
We derive a canonical form of the metric for such solutions
that effectively reduces the Einstein equations
to a differential equation on an axisymmetric $D-2$ by $D-2$ matrix field
living in three-dimensional flat space (apart from a subclass of solutions
that instead reduce to a set of equations on a $D-2$ by $D-2$
matrix field living in two-dimensional flat space).
This generalizes the Papapetrou form of the metric
for stationary and axisymmetric solutions in four dimensions,
and furthermore generalizes the work on Weyl solutions in four and higher
dimensions.
We analyze then
the sources for the solutions, which are in the form of thin rods
along a line
in the three-dimensional flat space that the
matrix field can be seen to live in.
As examples of stationary and axisymmetric solutions, we study
the five-dimensional rotating black hole
and the rotating black ring,
write the metrics in the canonical form and
analyze the structure of the rods for each solution.

%\vskip 0.5cm
%\leftline{\timestamp}

\end{titlepage}

\pagestyle{empty}
\small
\tableofcontents
\normalsize
\newpage
\pagestyle{plain}
\setcounter{page}{1}

%%%%%%%%%%%%%%%%%%%%%%%%%%%%%%%%%%%%%%%%%%%%%%%%%%%%%%%%%%%%%%
\section{Introduction}

Black holes in four-dimensional General Relativity have been the subject
of intense research for several decades.
One of the most important results on four-dimensional black holes
in pure gravity, i.e. gravity without matter,
is the uniqueness theorem stating that the rotating
black hole solution of Kerr \cite{Kerr:1963ud}
is the unique solution for given
mass and angular momentum
\cite{Israel:1967wq,Carter:1971,Hawking:1972vc,Robinson:1975}.
This shows that the phase structure of black holes in four
dimensions is very simple: Only one phase is available.

In recent years, attention have turned to the study of black
holes in higher-dimensional General Relativity.
It is by now clear that the phase structure of black holes
is much more complicated when having more than four dimensions.
For five-dimensional asymptotically flat black hole solutions,
it was discovered by Emparan and Reall
in \cite{Emparan:2001wn} that in addition to the
Myers-Perry rotating black hole solution \cite{Myers:1986un},
which has horizon topology $S^3$, there exists also a
rotating black ring solution
with horizon topology $S^2 \times S^1$.
This means that for a given mass and angular momentum one can
have as many as three different available phases,
for five-dimensional
asymptotically flat solutions of pure gravity.
For pure gravity solutions asymptoting to
Minkowski-space times a circle $\CM^d \times S^1$,
one has an even richer phase structure, involving
phases with different horizon topologies and also phases
with Kaluza-Klein bubbles.%
\footnote{See \cite{Elvang:2004iz,Elvang:2004ny} and
references therein.}

The complicated and rich phase structure of black holes in
higher dimensions makes
it desirable to develop new tools to find exact solutions.
We focus in this paper on a particular class of solutions:
Stationary and axisymmetric
solutions of the vacuum Einstein equations
in higher-dimensional General Relativity, i.e. in pure gravity.
These solutions have $D-2$ commuting Killing vector fields
where $D$ is the dimension of the space-time.
In four dimensions, this class of solutions includes
the Kerr black hole \cite{Kerr:1963ud}, while in five dimensions
both the rotating black hole with horizon topology $S^3$
\cite{Myers:1986un}
and the rotating black ring with horizon topology $S^2 \times S^1$
\cite{Emparan:2001wn}
are in this class.

We find in this paper a canonical form of the metric
for stationary and axisymmetric
solutions of the vacuum Einstein equations
in higher-dimensional General Relativity.
With the metric in the canonical form, the
Einstein equations takes a remarkably simple
form: They reduce effectively to
a differential equation that can be seen as
an axisymmetric $D-2$ by $D-2$ matrix field $G$
living in a three-dimensional flat space,
apart from a subclass of solutions
that instead reduce to a set of equations on a $D-2$ by $D-2$
matrix field living in two-dimensional flat space.

We analyze the general structure of such solutions.
In the three-dimensional space that $G$ can be seen to live in
the sources for $G$ are in the form of thin rods along a line.
We examine the general structure of the rods
that constitute the sources of a given solution.
We furthermore identify the asymptotic behavior
of asymptotically flat solutions in four and five dimensions.

As examples of stationary and axisymmetric solutions, we consider the
five-dimensional rotating black hole with horizon topology $S^3$
and the black ring with horizon topology $S^2 \times S^1$.
We write down the metric in the canonical coordinates and analyze
their rod-structure, i.e. the structure of their sources.

In four dimensions, the canonical form of the metric that we find
for stationary and axisymmetric solutions is equivalent
to the so-called {\sl Papapetrou form}
for the metric \cite{Papapetrou:1953,Papapetrou:1966}.
Papapetrou found that, under certain conditions,
the metric of
four-dimensional stationary and axisymmetric pure gravity solutions
can be written in the form%
\footnote{See also \cite{Stephani:2003,Wald:1984rg,Chandrasekhar:1992,Heusler:1996}.}
\begin{equation}
\label{papmet}
ds^2 = - e^{2U} ( dt + A d\phi )^2 + e^{-2U} r^2 d\phi^2
+ e^{2\nu} ( dr^2 + dz^2 )  \ .
\end{equation}
The functions $U(r,z)$, $A(r,z)$ are solutions of
\begin{equation}
\label{UAeqs}
\left( \partial_r^2 + \frac{1}{r} \partial_r + \partial_z^2 \right) U
= - \frac{e^{4U}}{2r^2} \left[ (\partial_r A)^2 + (\partial_z A)^2 \right]
\spa
\partial_r \left( \frac{e^{4U}}{r} \partial_r A \right)
+ \partial_z \left( \frac{e^{4U}}{r} \partial_z A \right) = 0 \ ,
\end{equation}
and the function $\nu(r,z)$ is a solution of
\begin{equation}
\label{papnu}
\begin{array}{c} \ds
\partial_r \nu = - \partial_r U + r \left[ (\partial_r U)^2
- (\partial_z U)^2 \right] - \frac{e^{4U}}{4r} \left[ (\partial_r A)^2
- (\partial_z A)^2 \right] \ ,
\\[3mm] \ds
\partial_z \nu = - \partial_z U + 2r \partial_r U \partial_z U
- \frac{e^{4U}}{2r} \partial_r A \partial_z A \ .
\end{array}
\end{equation}
Here $\partial/\partial t$ and $\partial / \partial \phi$
are the two Killing vector fields.
Since the equations \eqref{papnu}
for $\nu$ are integrable, one can solve the Einstein
equations by first finding $U$ and $A$ that solves
\eqref{UAeqs}, and then a $\nu$ can be found
that solves \eqref{papnu}.

The canonical form of the metric for stationary and axisymmetric solutions
that we find in this paper
is a generalization of the Papapetrou form \eqref{papmet}
of the metric for four-dimensional solutions.
Moreover, the simplified form of the Einstein equations that we find
generalizes
the equations \eqref{UAeqs}-\eqref{papnu} for four dimensions.

For the special case when all the $D-2$ Killing vector fields
are orthogonal to each other,
the canonical form of the metric that we find in this paper
is equivalent to the form of the so-called
generalized Weyl solutions of Emparan and Reall \cite{Emparan:2001wk}.%
\footnote{See \cite{Elvang:2004iz} for a brief review of
generalized Weyl solutions. See furthermore
\cite{Charmousis:2003wm} for work on extending the generalized
Weyl solutions of \cite{Emparan:2001wk} to space-times with a
cosmological constant.} In Ref.~\cite{Emparan:2001wk} it is shown
that, under certain conditions, the metric for $D$-dimensional
pure gravity solutions with $D-2$ commuting orthogonal Killing
vector fields can be written in the form
\begin{equation}
\label{GWmet}
ds^2 = - e^{2U_1} dt^2 + \sum_{i=2}^{D-2} e^{2U_i} (dx^i)^2
+ e^{2\nu} (dr^2+dz^2)
\spa
\sum_{i=1}^{D-2} U_i = \log r \ ,
\end{equation}
with $t=x^1$. The functions $U_i(r,z)$ are solutions of the
three-dimensional Laplace equations
\begin{equation}
\label{Ulap}
\left( \partial_r^2 + \frac{1}{r} \partial_r + \partial_z^2 \right) U_i
=0 \ ,
\end{equation}
for $i=1,...,D-2$, while $\nu(r,z)$ is a solution of
\begin{equation}
\label{GWnu}
\partial_r \nu = - \frac{1}{2r }
+ \frac{r}{2} \sum_{i=1}^{D-2} \left[ (\partial_r U_i)
- (\partial_z U_i)^2 \right]
\spa
\partial_z \nu = r \sum_{i=1}^{D-2} \partial_r U_i \partial_z U_i \ .
\end{equation}
Here $\partial / \partial x^i$, $i=1,...,D-2$, are the $D-2$
orthogonal Killing vector fields. Solutions with metric
\eqref{GWmet} with $U_i$ and $\nu$ obeying
\eqref{Ulap}-\eqref{GWnu} are called {\sl generalized Weyl
solutions}.

We see that using the form of the metric \eqref{GWmet}
for solutions with $D-2$ commuting orthogonal
Killing vector fields, solving the Einstein equations effectively
reduces to the task of solving $D-3$ free Laplace equations
on a three-dimensional flat space. This is due to the fact
that the equations \eqref{GWnu} for $\nu$ are integrable, so
that one can find a $\nu$ solving \eqref{GWnu} given any solution for
$U_i$, $i=1,...,D-2$.

It is important to remark that
the method of generalized Weyl solutions generalizes Weyl's work on
four-dimensional static and axisymmetric solutions \cite{Weyl:1917}.
Moreover, one also obtains Weyl's form of the metric for
four-dimensional static and axisymmetric solutions
by setting $A=0$ in Papapetrou form \eqref{papmet}.
This is consistent with the fact that Eqs.~\eqref{papmet}-\eqref{papnu}
becomes equivalent to Eqs.~\eqref{GWmet}-\eqref{GWnu} for $D=4$
when $A=0$,
with $U_1 = U$ and $U_2 = \log r - U$.

Eqs.~\eqref{Ulap} can be seen
as free Laplace equations for axisymmetric potentials living in
a three-dimensional flat space. Solutions are then
build up from thin rods located at the line $r=0$ in the three-dimensional
space, with a given
rod being a source for one of the $D-2$ potentials $U_i$ \cite{Emparan:2001wk}.
In this paper we generalize the concept of rods to the
more general class of stationary and axisymmetric solutions,
i.e. solutions for which the Killing vector
fields are not necessarily orthogonal.
One of the new features is that for a given rod we can associate a direction
in the $(D-2)$-dimensional vector space spanned by the Killing
vector fields.
Solutions for which the directions of the rods are not orthogonal
to each other
are then also solutions where the Killing vector fields are not
orthogonal to each other.

The outline of this paper is as follows:
In Section \ref{secstataxi} we derive a canonical form of the metric
for stationary and axisymmetric pure gravity solutions.
Using this, we find a simplified version of the Einstein equations,
effectively reducing them to an equation on an
axisymmetric $D-2$ by $D-2$
matrix field $G$ living in flat three-dimensional space.
Some of the details of the derivation are placed in
the Appendices \ref{appdetG}, \ref{2Dmet} and \ref{appricci}.
In Appendix \ref{appdetconst} we consider a special subclass of
solutions that has the matrix field $G$ living in two-dimensional
flat space.
In Appendix \ref{appprop} we explore further the equation
for the matrix field $G$.

In Section \ref{s:nearrod} we consider the behavior of the
matrix field $G$ near the $r=0$ line in the flat three-dimensional
space that $G$ lives in. The sources for $G$ lives on the $r=0$ line
in the form of rods. We analyze the general structure of these rods.
See also Appendix \ref{appsing}.

In Section \ref{s:asymp} consider the asymptotic region, and we
find out how to read off the asymptotic quantities for solutions
that asymptotic to four-dimensional or five-dimensional Minkowski-space.

In Section \ref{s:rotBH} and \ref{s:BRsol} we write down the
metrics for the
five-dimensional rotating black hole of Myers and Perry and
the rotating black ring of Emperan and Reall in the canonical form.
We furthermore analyze the rod-structure for these solutions.
For the rotating black hole solutions, we make use
of Appendix \ref{a:prolate} on prolate spherical coordinates,
while for the black ring solutions we make use of Appendix \ref{s:cmet}
which considers C-metric coordinates and how to transform these
to the canonical coordinates of this paper.

In Section \ref{s:concl} we have the conclusions.

%%%%%%%%%%%%%%%%%%%%%%%%%%%%%%%%%%%%%%%%%%%%%%%%%%%%%%%%%%%%%%
\section{Stationary and axisymmetric solutions}
\label{secstataxi}

In this section we show that finding
stationary and axisymmetric solutions of General Relativity
in $D$ dimensions without matter (i.e. pure gravity)
can be reduced to finding solutions of a
differential equation on an axisymmetric $D-2$ by $D-2$
matrix field in
flat three-dimensional Euclidean space.
As part of this, we find a particularly simple form of the metric
for such solutions.

With respect to four-dimensional General Relativity, the
results of this section generalizes the work of Papapetrou
on stationary and axisymmetric metrics in four dimensions
\cite{Papapetrou:1953,Papapetrou:1966}
(see Eqs.~\eqref{papmet}-\eqref{papnu} in the Introduction),
which again is a generalization of the work of Weyl
on static and axisymmetric metrics \cite{Weyl:1917}.
In higher-dimensional General Relativity, the results of this section
generalizes the work of Emparan and Reall on
metrics with $D-2$ orthogonal commuting Killing vector fields
\cite{Emparan:2001wk}
(see Eqs.~\eqref{GWmet}-\eqref{GWnu} in the Introduction).
We comment in more detail
on the connection to previous work in the following.
Finally, we note that the derivation of this section follows similar
lines as that of Wald's derivation in \cite{Wald:1984rg}
for four-dimensional stationary and axisymmetric metrics.

%%%%%%%%%%%%%%%%%%%%%%%%%%%%%%%%%%%%%%
\subsection{Deriving canonical form of metric and the Einstein equations}

%%%%%%%%%%%%%%%%%%%%%
\subsubsection*{Formulation of problem}

In this section we study $D$-dimensional manifolds which have $D-2$
commuting linearly independent Killing vector fields
$V_{(i)}$, $i=1,...,D-2$.
With Lorentzian signature this corresponds to what we in this paper
call {\sl stationary and axisymmetric} space-times, where
the term ``stationary'' means that one of our Killing vector fields are
time-like, while the $D-3$ space-like Killing vector fields
gives what we call ``axisymmetry'' of the space-time.%
\footnote{One can use our results for
null Killing vector fields, but we will not elaborate on that case
in this paper.}
That the Killing vector fields $V_{(i)}$, $i=1,...,D-2$, commute means that
\begin{equation}
\label{VVcom}
[ V_{(i)} , V_{(j)} ] = 0 \ ,
\end{equation}
for $i,j=1,...,D-2$. We see that the Killing vector fields
generate a $(D-2)$-dimensional abelian group.

We restrict moreover ourselves to consider solutions of $D$-dimensional
General Relativity without matter, i.e. we consider metrics that
solve the vacuum Einstein equations
\begin{equation}
\label{Riccifl}
R_{\mu \nu} = 0 \ .
\end{equation}
In the following we find a canonical form of this class of metrics
and we find furthermore a reduced form of the Einstein equations
\eqref{Riccifl}.

%%%%%%%%%%%%%%%%%%%%%%%%%%%%%%%%%%%%
\subsubsection*{Finding two-dimensional orthogonal subspaces}

Consider first a general $D$-dimensional space-time with
$D-2$ commuting Killing vector fields $V_{(i)}$, $i=1,...,D-2$.
From the fact that the Killing vector fields
are commuting, as expressed in Eq.~\eqref{VVcom},
we get that we can find coordinates $x^i$, $i=1,...,D-2$,
and $u^a$, $a=1,2$, so that
\begin{equation}
V_{(i)} = \frac{\partial}{\partial x^i} \ ,
\end{equation}
for $i=1,...,D-2$. Clearly, this means that
the metric components in this coordinate
system only depends on $u^1$ and $u^2$.

We need now the theorem \cite{Wald:1984rg,Emparan:2001wk}:
\begin{theorem}
\label{theosubint}
Let $V_{(i)}$, $i=1,...,D-2$, be $D-2$ commuting Killing vector fields
such that:
\begin{itemize}
\item[(1)] The tensor
$V_{(1)}^{[\mu_1} V_{(2)}^{\mu_1} \cdots V_{(D-2)}^{\mu_{D-2}}
D^\nu V_{(i)}^{\rho]}$ vanishes at at least one point of the space-time
for a given $i=1,...,D-2$.
\item[(2)] The tensor $V^\nu_{(i)} R_\nu^{\ [\rho}
V_{(1)}^{\mu_1} V_{(2)}^{\mu_1} \cdots V_{(D-2)}^{\mu_{D-2}]} = 0$
for all $i=1,...,D-2$.
\end{itemize}
Then the two-planes orthogonal to the Killing vector
fields $V_{(i)}$, $i=1,...,D-2$, are integrable.
\squ
\end{theorem}

This theorem is stated and proven in four dimensions in \cite{Wald:1984rg}
using Frobenius theorem on integrable submanifolds.
Emparan and Reall generalized it to higher-dimensional
manifolds in \cite{Emparan:2001wk}.

Assume now that the two conditions in Theorem \ref{theosubint} are obeyed.
That the two-planes orthogonal to the Killing vector
fields $V_{(i)}$, $i=1,...,D-2$, are integrable means that
 for any given point of our $D$-dimensional
manifold we have a two-dimensional submanifold that includes this point and
moreover have the property that for any point of the submanifold
the two-dimensional
tangent-space is orthogonal to all of the Killing vector
fields.
By choosing coordinates on one of these two-dimensional submanifolds and
dragging them along the integral curves of our Killing vector fields,
we can find two coordinates $y^1$ and $y^2$ for our $D$-dimensional
manifold so that
$\frac{\partial}{\partial x^i}$ is orthogonal
to $\frac{\partial}{\partial y^a}$
everywhere for all $i=1,...,D-2$ and $a=1,2$.
This means the metric takes the form
\begin{equation}
\label{met1}
ds^2 = \sum_{i,j=1}^{D-2} G_{ij} dx^i dx^j
+ \sum_{a,b=1}^2 \hat{g}_{ab} dy^a dy^b \ ,
\end{equation}
where $G_{ij}$ and $\hat{g}_{ab}$ only depends on $y^1$ and $y^2$.

From now on we restrict ourselves to solutions solving
the vacuum Einstein equations \eqref{Riccifl}.
This ensures immediately that Condition (2) in Theorem \ref{theosubint}
is obeyed. We assume furthermore that Condition (1)
in Theorem \ref{theosubint} is obeyed. Condition (1) can for example
be argued to hold if one of the Killing vector fields is an angle, since
then it is zero on the axis of rotation.
This means for instance that
solutions asymptoting to
Minkowski-space $\CM^D$ for $D=4,5$ obeys Condition (1) since they
have angles in them. Clearly, the same is true for solutions asymptoting to
$\CM^{D-p} \times T^p$ for $D-p=4,5$.

%%%%%%%%%%%%%%%%%%%%%%%%%%%%%
\subsubsection*{The $r$ and $z$ coordinates}

Define now the function $r(y^1,y^2)$ as
\begin{equation}
r = \sqrt{ | \det ( G_{ij} ) | } \ .
\end{equation}
In Appendix \ref{appdetconst} we treat the case in which $\det (G_{ij})$ is
constant, giving rise to a special class of solutions.
Instead, we assume here and in the following
that $r(y^1,y^2)$ is not a constant function.
From Appendix \ref{appdetG} we get then that
$(\frac{\partial r}{\partial{y^1}},\frac{\partial r}{\partial {y^2}})
\neq (0,0)$ except in isolated points.
We can then use the result of Appendix \ref{2Dmet} that we can
find a coordinate $z(y^1,y^2)$, along with two functions
$\nu(y^1,y^2)$ and $\Lambda(y^1,y^2)$, so that
\begin{equation}
\sum_{a,b=1}^2 \hat{g}_{ab} dy^a dy^b = e^{2\nu}
 ( dr^2 + \Lambda \, dz^2 ) \ .
\end{equation}
Therefore, the full metric takes the form
\begin{equation}
\label{met2}
ds^2 = \sum_{i,j=1}^{D-2} G_{ij} dx^i dx^j +
e^{2\nu} ( dr^2 + \Lambda \, dz^2 ) \ ,
\end{equation}
where $\nu(r,z)$ and $\Lambda(r,z)$ are functions of $r$ and $z$.

From Appendix \ref{appricci}, where part of the Ricci tensor
for the metric \eqref{met2} is computed, we have from Eq.~\eqref{trGR}
\begin{equation}
\sum_{i,j=1}^{D-2} G^{ij} R_{ij}
= - \frac{\partial_r \Lambda}{2 e^{2\nu} \Lambda r} \ .
\end{equation}
Since our solution should fulfil the
vacuum Einstein equations $R_{\mu \nu} = 0$, this means that
\begin{equation}
\partial_r \Lambda = 0 \ .
\end{equation}
This gives that $\Lambda = \Lambda (z)$. Since
we preserve the form of the metric \eqref{met2}
under a transformation $z' = f(z)$ we can therefore
set $\Lambda(z)=1$ by a coordinate transformation of $z$ alone.
Thus, we can define the $z$-coordinate by demanding $\Lambda=1$.
This fixes $z$ up to transformations
$z \rightarrow z + \mbox{constant}$.

%%%%%%%%%%%%%%%%%%%%%%%%%%%%%%%%%%%%%%%%%%%%
\subsubsection*{Canonical form of metric}

In conclusion, we have shown that for any Ricci-flat
space-time with $D-2$ commuting Killing vector fields $V_{(i)}$,
$i=1,...,D-2$, obeying Condition (1) of
Theorem \ref{theosubint},
we can find a coordinate system $(x^1,...,x^{D-2},r,z)$ such that
$V_{(i)} = \frac{\partial}{\partial x^i}$ and such that
the metric takes the {\sl canonical form}
\begin{equation}
\label{themet}
ds^2 = \sum_{i,j=1}^{D-2} G_{ij} dx^i dx^j +
e^{2\nu} ( dr^2 + dz^2 ) \ ,
\end{equation}
with
\begin{equation}
\label{rdetG}
r = \sqrt{ | \det ( G_{ij} ) | } \ ,
\end{equation}
where $G_{ij} (r,z) $ and $\nu (r,z)$ are functions only of $r$ and $z$.
In addition to the assumption that the Killing vector fields should
obey Condition (1) of Theorem \ref{theosubint} we also assume here
that $\det (G_{ij})$ is not constant on our space-time.
The situation in which $\det (G_{ij})$ is constant is instead treated in
Appendix \ref{appdetconst}.

%%%%%%%%%%%%%%%%%%%%%%%%%%%%%%%%%%%%%%%%%%%%%%%%
\subsubsection*{The Einstein equations}

We now consider the vacuum Einstein
equations $R_{\mu \nu} = 0$ for the metric \eqref{themet} with the
constraint \eqref{rdetG} using the computed Ricci tensor
\eqref{RicciComp} in Appendix \ref{appricci}.

Considering the $R_{ij}=0$ equations
we see from \eqref{RicciComp} that the equations for $G_{ij}$ are
\begin{equation}
\label{Gijeqs}
\left( \partial_r^2 + \frac{1}{r} \partial_r + \partial_z^2 \right) G_{ij}
= \sum_{k,l=1}^{D-2}
 G^{kl} \partial_r G_{ki} \partial_r G_{lj}
+ \sum_{k,l=1}^{D-2}
G^{kl} \partial_z G_{ki} \partial_z G_{lj}  \ .
\end{equation}
Considering the $R_{rr}-R_{zz}=0$ and $R_{rz}=0$ equations
we see from \eqref{RicciComp} that the equations for $\nu$ are
\begin{equation}
\label{nueqs}
\begin{array}{c} \ds
\partial_r \nu = - \frac{1}{2r}
+ \frac{r}{8} \sum_{i,j,k,l=1}^{D-2}  G^{ij}
G^{kl} \partial_r G_{ik} \partial_r G_{jl}
- \frac{r}{8} \sum_{i,j,k,l=1}^{D-2}  G^{ij}
G^{kl} \partial_z G_{ik} \partial_z G_{jl}  \ ,
\\[4mm] \ds
\partial_z \nu = \frac{r}{4} \sum_{i,j,k,l=1}^{D-2}  G^{ij}
G^{kl} \partial_r G_{ik} \partial_z G_{jl}  \ .
\end{array}
\end{equation}
Using now \eqref{nueqs} together
with \eqref{Gijeqs} one can check
that the integrability condition
$\partial_z \partial_r \nu = \partial_r \partial_z \nu$ on $\nu(r,z)$ is
obeyed.
Thus, for a given solution $G_{ij}(r,z)$ of \eqref{Gijeqs}
the equations \eqref{nueqs}
can be integrated to give $\nu(r,z)$.

Finally, there is the remaining non-trivial equation
$R_{rr} + R_{zz} = 0$
coming from the Einstein equations.
The explicit expression for this
equation is easily found using \eqref{RicciComp} and
is seen to involve second derivatives of $\nu$.
Since $G_{ij}(r,z)$ and $\nu(r,z)$
already are determined by \eqref{Gijeqs}-\eqref{nueqs} it
needs to be checked that $R_{rr} + R_{zz} = 0$ is consistent
with \eqref{Gijeqs}-\eqref{nueqs}.
This can be checked by finding $\partial_r^2 \nu + \partial_z^2 \nu$
from \eqref{nueqs}. Inserting the result
into $R_{rr} + R_{zz}$ from \eqref{RicciComp}
this is seen to be zero using \eqref{Gijeqs}.

Therefore, we have shown that one can find solutions
of the vacuum Einstein equations for the canonical
form for the metric \eqref{themet}-\eqref{rdetG}
by finding a $G_{ij} (r,z)$
that satisfies \eqref{Gijeqs}. Then, subsequently one can always
find a function $\nu(r,z)$ that satisfies \eqref{nueqs},
and thereby we have a complete solution satisfying all
the Einstein equations.

%%%%%%%%%%%%%%%%%%%%%%%%%%%%%%%%%%%%%%%%%
\subsubsection*{Reduction to Papapetrou form and generalized
Weyl solutions}

We show here that the canonical form of the metric
\eqref{themet}-\eqref{rdetG}, along with the
form of the Einstein equations \eqref{Gijeqs}-\eqref{nueqs},
reduces to the previously
known cases.

We first consider the Papapetrou form \eqref{papmet} for
four dimensional stationary and axisymmetric solutions
\cite{Papapetrou:1953,Papapetrou:1966}, with the Einstein
equations in the form \eqref{UAeqs}-\eqref{papnu}.
Setting $D=4$, we see that by setting $G_{11} = - e^{2U}$,
$G_{12} = - e^{2U} A$ and
$G_{22} = e^{-2U} ( r^2 - A^2 e^{4U} )$ with $x^1=t$ and
$x^2=\phi$, we get the
Papapetrou form \eqref{papmet} from \eqref{themet}.
Furthermore, we see that \eqref{Gijeqs}-\eqref{nueqs}
reduce to \eqref{UAeqs}-\eqref{papnu}.

Consider now
instead the generalized Weyl solutions of \cite{Emparan:2001wk}
which have $D-2$ orthogonal commuting Killing vector fields.
These have metric \eqref{GWmet}, and the Einstein equations
are in the form \eqref{Ulap}-\eqref{GWnu}.
We see that setting $G_{11} = - e^{2U_1}$ and
$G_{ii} = e^{2U_i}$ for $i=2,..,D-2$, we get the metric
\eqref{GWmet} from the canonical form \eqref{themet}.
$\det G = - r^2$ gives then $\sum_{i=1}^{D-2} U_i = \log r$.
For the Einstein equations, it is easily seen
that \eqref{Gijeqs}-\eqref{nueqs} reduces to
\eqref{Ulap}-\eqref{GWnu} (see also Appendix \ref{appprop}).
Thus, the canonical form \eqref{themet}-\eqref{rdetG} correctly reduce
to the generalized Weyl solutions.

%%%%%%%%%%%%%%%%%%%%%%%%%%%%%%%%%%%%%%%%%%%%%%%%%%%%%%%%%%
\subsection{Compact notation for the equations for $G_{ij}(r,z)$}

We have derived above that the metric of $D$-dimensional manifolds with
$D-2$ commuting Killing vector fields obeying the vacuum
Einstein equations
can be written in the canonical form
\eqref{themet}-\eqref{rdetG}.
Moreover, the vacuum Einstein equations reduce to
\eqref{Gijeqs}-\eqref{nueqs}.
We now show that we can
write the equations for $G_{ij} (r,z)$ in a more compact form.
This is highly useful for analysis of these equations.

For a given $r$ and $z$
we can view $G_{ij}$ as a $D-2$ times $D-2$ real symmetric
matrix, with $G^{ij}$ as its inverse.
In this way we can write \eqref{Gijeqs} in matrix notation as
\begin{equation}
G^{-1} \left( \partial_r^2 + \frac{1}{r} \partial_r + \partial_z^2 \right) G
= (G^{-1} \partial_r G )^2 + (G^{-1} \partial_z G )^2  \ ,
\end{equation}
with the constraint $| \det G | = r^2$ coming from \eqref{rdetG}.

We can make a further formal rewriting of \eqref{Gijeqs} by
recognizing that the derivatives respects the symmetries
of a flat three-dimensional Euclidean space with metric
\begin{equation}
\label{unmet}
dr^2 + r^2 d\gamma^2 + dz^2 \ .
\end{equation}
Here $\gamma$ is an angular coordinate of period $2\pi$.%
\footnote{It is important to remark that
$\gamma$ is not an actual physical variable for the solution
\eqref{themet}, but rather an auxilirary coordinate that is
useful for understanding the structure of Eqs.~\eqref{Gijeqs}.}
Therefore, if we define $\grad$ to be the gradiant in
three-dimensional flat Euclidean space, we can write \eqref{Gijeqs} as
\begin{equation}
\label{formalG}
G^{-1} \grad^2 G = ( G^{-1} \grad G)^2  \ .
\end{equation}
Thus, by finding the axisymmetric solutions of the
differential matrix equation \eqref{formalG} in
three-dimensional flat Euclidean space, that obey the constraint
$| \det G | = r^2$, we can find all stationary
and axisymmetric solutions of the vacuum Einstein equations
in $D$ dimensions.%
\footnote{To be precise, we mean all solutions
for which the $D-2$ Killing vector fields
obey Condition (1) of Theorem \ref{theosubint} and
for which $\det ( G_{ij} )$ is a non-constant function on
our $D$-dimensional manifold.}

We explore some of the mathematical properties
of Eq.~\eqref{formalG} in Appendix \ref{appprop}.
Here the compact form \eqref{formalG} of \eqref{Gijeqs}
prove highly useful.

%%%%%%%%%%%%%%%%%%%%%%%%%%%%%%%%%%%%%%%%%%%%%%%%%%%%%%%%%%%%%%
\section{Behavior of solutions near $r=0$}
\label{s:nearrod}

In Section \ref{secstataxi} we derived
the canonical form of the metric \eqref{themet}-\eqref{rdetG},
along with the corresponding EOMs \eqref{Gijeqs}-\eqref{nueqs},
for stationary and axisymmetric solutions of the vacuum Einstein equations.
In this section we consider the behavior of such solutions
close to $r=0$.

%%%%%%%%%%%%%%%%%%%%%%%%%%%%%%%%%%%%%%%%%%%
\subsection{Behavior of $G(r,z)$ near $r=0$}
\label{s:rods}

We describe first how the $z$-axis at $r=0$
is divided into intervals, called rods,
according to the dimension of the kernel of $G$ at $r=0$.
We find then coordinates in which a solution simplifies near a rod,
making it possible to describe the solution in detail near the rod.
We use this to define the rod-structure of a solution
in Section \ref{s:genrod}.

%%%%%%%%%%%%%%%%%%%%%%%
\subsubsection*{Dividing the $z$-axis into rods}

Consider a given solution $G(r,z)$.
$G(r,z)$ is required to be continuous.
Since $|\det G| = r^2$ we see that the product of the eigenvalues
of $G(r,z)$ goes to zero for $r \rightarrow 0$.
Therefore, we have that the eigenvalues of $G(0,z)$, which
all are real since $G(0,z)$ is symmetric, include the eigenvalue zero
for a given $z$.
This means that the dimension of the kernel of $G(0,z)$ is greater
than or equal to one for any $z$. We can write this
more compactly as $\dim ( \ker ( G(0,z) )) \geq 1$.

A necessary condition for a regular
solution is that
precisely one eigenvalue of $G(0,z)$ is zero for a
given $z$, except in isolated points.
This statement is explained in Appendix \ref{appsing} where we argue that
if we have more than one eigenvalue going to zero as
$r \rightarrow 0$, for a given $z$, we have a curvature
singularity at that point.
Therefore, in the following we consider only solutions which, for a given
$z$, only have one eigenvalue going to zero for $r\rightarrow 0$,
except at isolated values of $z$.
Written compactly, this means $\dim ( \ker ( G(0,z)) = 1$,
except at isolated values of $z$.
Denote now these isolated values of $z$ as $a_1,a_2,...,a_N$,
with $a_1 < a_2 < \cdots < a_N$.

We see now that we divided the $z$-axis into the $N+1$ intervals
$[-\infty,a_1]$, $[a_1,a_2]$, ..., $[a_{N-1},a_N]$ and $[a_N,\infty]$.%
\footnote{Note that it is possible to have an infinite number of intervals.}
We call these $N+1$ intervals the {\sl rods} of the solution.

One can easily check that the above definition of rods
reduces to the definition of
\cite{Emparan:2001wk} for the special case of
generalized Weyl solutions, i.e. with $D-2$ orthogonal Killing vector fields.

%%%%%%%%%%%%%%%%%%%%%%%
\subsubsection*{Behavior of $G(r,z)$ near a rod}

In Section \ref{secstataxi} we found that $G(r,z)$ should solve
 the equation
$G^{-1} \grad^2 G = (G^{-1} \grad G)^2$ with the constraint
that $|\det G | = r^2$.
However, this breaks down as $r\rightarrow 0$, because
for $r=0$ we have that
$\det G= 0$ so $G$ is not invertible anymore.
The reason for this is that we have sources added to the equation
$G^{-1} \grad^2 G = (G^{-1} \grad G)^2$ at $r=0$.
The sources corresponds precisely to the rods defined above,
i.e. the intervals with $\dim ( \ker ( G(r,z) ))=1$.
Moreover, if we view the solution $G(r,z)$ as a matrix-valued field in the
unphysical three-dimensional flat Euclidean space with metric
\eqref{unmet}, a rod is really a source in the form of a rod
of zero thickness in this unphysical space.
In the following we examine in detail the equation
$G^{-1} \grad^2 G = (G^{-1} \grad G)^2$ near a rod in order to
describe more precisely the behavior of $G(r,z)$ near a rod.

Consider a solution $G(r,z)$ and a given rod $[z_1,z_2]$.
Consider furthermore a given value of $z=z_*$ obeying $z_1 < z_* < z_2$.
Since $G(0,z_*)$ is a symmetric real matrix we can
diagonalize it using an orthogonal matrix $\Lambda_*$
such that $\Lambda_*^T G(0,z_*) \Lambda_*$
is diagonal. Furthermore, since $G(0,z_*)$ has precisely
one zero eigenvalue, we can choose $\Lambda_*$ so that
$( \Lambda_*^T G(0,z_*) \Lambda_* )_{11} = 0$.

Define now $\tilde{G} (r,z) = \Lambda_*^T G(r,z) \Lambda_*$.
Clearly, $\tilde{G} (r,z)$ is a solution of \eqref{Gijeqs}
by Lemma \ref{lemmult}, and furthermore
$\det \tilde{G} = \det G = \pm r^2 $.
Note that all entries of $\tilde{G}(r,z_*)$
are of order $\CO (r^2)$ for $r \rightarrow 0$,
expect the entries
$\tilde{G}_{ii} (r,z_*)$, $i=2,...,D-2$, which are finite and non-zero.

Since $\tilde{G}^{ij} (0,z_*)$ is not well-defined we need
to consider the limit $\tilde{G}^{ij} (r,z_*) $
for $r \rightarrow 0$ carefully.
To this end, define
the $D-2$ by $D-2$ matrix-valued function $M(r,z)$ by
\begin{equation}
M_{11} = \frac{\tilde{G}_{11} }{r^2}
\spa
M_{1i} = \frac{\tilde{G}_{1i} }{r}
\spa
M_{ij} = \tilde{G}_{ij}
\spa
i,j = 2,...,D-2
\end{equation}
for any $(r,z)$. We see that this corresponds to a rescaling
$x_{\rm new}^1 = r x_{\rm old}^1$.
Clearly, we have that $M(0,z_*)$ is diagonal,
with non-zero eigenvalues.
Moreover, we have that
$\tilde{G}^{11} = M^{11} / r^2$, $\tilde{G}^{1i} = M^{1i} / r$
and $\tilde{G}^{ij} = M^{ij}$, $i,j = 2,...,D-2$.
Since $M_{1i}$, $i=2,..,D-2$, are of order
$\CO(r)$, we have that $M^{1i}$, $i=2,..,D-2$, are of order
$\CO(r)$, and therefore that $\tilde{G}^{1i} (r,z_*)$, $i=2,..,D-2$,
stay finite (or goes to zero) in the limit $r \rightarrow 0$.
Also, $\tilde{G}^{ij} (r,z_*) \rightarrow 0$ with $2 \leq i < j \leq D-2$
and
$\tilde{G}^{ii} (r,z_*) \rightarrow (\tilde{G}_{ii} (0,z_*))^{-1}$
with
$i=2,...,D-2$, while $\tilde{G}^{11} (r,z_*)$ is of order $1/r^2$
for $r \rightarrow 0$.

Consider now the equation $\tilde{G}^{-1} \grad^2 \tilde{G}
= (\tilde{G}^{-1} \grad \tilde{G})^2$ for
$z=z_*$ and $r\rightarrow 0$.
We have
\begin{equation}
\label{grad2G11}
\grad^2 \tilde{G}_{11} = \tilde{G}^{11} (\grad \tilde{G}_{11})^2
+ 2 \sum_{i=2}^{D-2} \tilde{G}^{1i} \grad \tilde{G}_{11} \cdot
\grad \tilde{G}_{1i}
+ \sum_{i=2}^{D-2} \tilde{G}^{ii} (\grad \tilde{G}_{1i})^2 \ ,
\end{equation}
up to terms that go to zero for $r \rightarrow 0$.
Since we just found that $\tilde{G}^{1i}$ and $\tilde{G}^{ii}$
are finite for $z=z_*$ and $r \rightarrow 0$ (with $i=2,...,D-2$) and
since we require that $\tilde{G}_{ij} (r,z)$ and its derivatives
are finite as a necessary condition for regular solutions,%
\footnote{Except in the endpoints of a rod where the derivatives
are not necessarily well-defined. Hence the condition $z_1 < z_* < z_2$.}
we see that the LHS and the second and third term
on the RHS of \eqref{grad2G11}
are finite for $z=z_*$ and $r\rightarrow 0$.
Since $\tilde{G}^{11} \rightarrow \infty$
we see therefore that we need $\grad \tilde{G}_{11} \rightarrow 0$
for $z=z_*$ and  $r \rightarrow 0$.
If we consider instead $\grad^2 \tilde{G}_{ii}$ we see similarly that
\begin{equation}
\grad^2 \tilde{G}_{ii} = \tilde{G}^{11} (\grad \tilde{G}_{1i})^2
+ \mbox{finite terms} \ ,
\end{equation}
so we get that $\grad \tilde{G}_{1i} \rightarrow 0$ for
$z=z_*$ and  $r \rightarrow 0$.
Thus, we have derived that
$\grad \tilde{G}_{1i} (0,z_*) = 0$ for $i=1,...,D-2$.
In particular, this implies that $\partial_z \tilde{G}_{1i} (0,z_*)=0$.
Therefore, since this works for any $z_* \in \, ] z_1, z_2 [$ we get the
following theorem:
\begin{theorem}
\label{theorod}
Consider a rod $[z_1,z_2]$ for a solution $G(r,z)$.
Then we can find an orthogonal matrix $\Lambda_*$ such that
the solution
$\tilde{G}(r,z) = \Lambda_*^T G(r,z) \Lambda_*$ has the property that
$\tilde{G}_{1i} (0,z) = 0$ for $i=1,...,D-2$ and $z \in [z_1,z_2]$.
\squ
\end{theorem}

We take now a closer look at the EOMs \eqref{Gijeqs}-\eqref{nueqs}
near $r=0$.
Consider a solution $G(r,z)$ and a particular rod $[z_1,z_2]$.
Using Theorem \ref{theorod} we always make a constant coordinate
transformation of the $x^i$ coordinates so that
$G(r,z)$ has the property that
$G_{1i} (0,z) = 0$ for $i=1,...,D-2$ and $z \in [z_1,z_2]$.
To leading order, we can therefore write
$G(r,z) = ( \pm a(z) r^2 ) \oplus A(z)$
for $r \rightarrow 0$ with $z_1 < z < z_2$
where $a(z)$ is a function of $z$ with $a(z) > 0$
for $z \in \, ]z_1,z_2[$
and $A(z)$ is a $D-3$ by $D-3$ matrix-valued
function of $z$.
Thus, $G_{11} = \pm a (z) r^2$ for $r \rightarrow 0$.
Note that $| \det (A (z)) | = 1/a(z) $.

If we consider Eqs.~\eqref{Gijeqs} we see that
$\grad^2 G_{11} = \pm 4a(z) + \CO(r)$
and that $G^{11} ( \partial_r G_{11} )^2 = \pm 4a(z) + \CO(r)$, so
this is consistent.
Considering $\partial_r \nu$ in  \eqref{nueqs} we see that since
$\partial_r G_{11} = \pm 2 a(z) r $ we have that
$\partial_r \nu = 0$ to leading order.
Considering instead $\partial_z \nu$ in \eqref{nueqs}, we get
\begin{equation}
\partial_z \nu = \frac{1}{2} \frac{a'}{a} + \CO(r) \ .
\end{equation}
Thus, to leading order for $r\rightarrow 0$ we have
$e^{2\nu} = c^2 a(z)$ where $c$ is a positive number.
Therefore, for $r \rightarrow 0$ with $z_1 < z < z_2$ the metric
\eqref{themet} has the form
\begin{equation}
\label{nearrod}
ds^2 = \sum_{i,j=2,...,D-2} A_{ij}(z) dx^i dx^j
+ a(z) \left[ \pm r^2 (dx^1)^2 + c^2 ( dr^2 + dz^2 ) \right] \ .
\end{equation}
This is the behavior of the canonical metric \eqref{themet}
near a rod.

Notice now that if $G_{11}/r^2$ is positive for $r \rightarrow 0$
the coordinate $x^1$ is space-like and
the metric \eqref{nearrod} has a conical singularity for $r \rightarrow 0$,
unless $x^1$ is periodic with period $2\pi c$.
For a regular solution,
this means that if we have a rod in a space-like direction
we have necessarily that this direction is periodic with the period
constrained from avoiding the conical singularity.

If $G_{11}/r^2$ is negative for $r \rightarrow 0$
the coordinate $x^1$ is time-like and we see that there is
a horizon at $r=0$ since $G_{11} = 0$.
Moreover, using the above argument for the space-like direction,
we see that the Wick rotated coordinate $i x^1$ must be periodic
with period $2\pi c$. This means that the horizon
has a temperature $T = 1 / (2\pi c)$ associated to it.

%%%%%%%%%%%%%%%%%%%%%%%%%%%%%%%%
\subsection{The rod-structure of a solution}
\label{s:genrod}

In this section we define what we mean by the {\sl rod-structure}
of a solution, and we discuss the general structure of rods,
in view of the considerations of Section \ref{s:rods}.

%%%%%%%%%%%%%%%%%%%%%%
\subsubsection*{Specifying the rod-structure of a solution}

Let a solution $G_{ij}(r,z)$ of
Eqs.~\eqref{rdetG}-\eqref{Gijeqs}
be given with $N+1$ rods which meet in the
$z$-values $a_1 < a_2 < ... < a_N$.
Introduce here the notation $a_0=-\infty$ and
$a_{N+1} =\infty$ in order to write the equations
below more compactly.
The solution $G(r,z)$ thus have the $N+1$ rods $[a_{k-1},a_k]$
with $k=1,...,N+1$.

Define now for the solution $G_{ij}(r,z)$
the $N+1$ vectors $v_{(k)}$ in $\R^{D-2}$, $k=1,...,N+1$,
by
\begin{equation}
\label{e:defv}
G(0,z) v_{(k)} = 0 \ \ \mbox{for} \ \ z \in [a_{k-1},a_{k}]
\ \ \ ,\ \ k=1,...,N+1 \ ,
\end{equation}
with $v_{(k)} \neq 0$ for all $k=1,...,N+1$.
In other words, $v_{(k)} \in \ker ( G(0,z) )$.
We call $v_{(k)}$ the {\sl direction} of the corresponding
rod $[a_{k-1},a_k]$.

We define then the {\sl rod-structure} of the
solution $G_{ij} (r,z)$ as the specification of the
rod intervals $[a_{k-1},a_k]$
plus the corresponding directions $v_{(k)}$, $k=1,...,N+1$.

Obviously, since $v_{(k)}$ is defined as an eigenvector,
it is only defined up to
a multiplicative factor (different from zero).
In other words, one should really regard $v_{(k)}$ as
an element of the real projective space $\R P^{D-3}$.

We now demonstrate that it follows from the considerations of
Section \ref{s:rods} that the above definition
of the rod-structure is meaningfull. This involves showing that
$v_{(k)}$ as defined in \eqref{e:defv} exists
and is unique, as   element in $\R P^{D-3}$

Observe first that
by Theorem \ref{theorod} we get for each of the $N+1$ rods
an orthogonal matrix $\Lambda_{(k)}$, $k=1,...,N+1$,
so that $(\Lambda_{(k)}^T G(0,z) \Lambda_{(k)})_{1i} = 0$
for $z \in [a_{k-1},a_{k}]$.
Define the unit vector $e = (1,0,...,0)$ in $\R^{D-2}$.
Note that from the above we have then that
$\Lambda_{(k)}^T G(0,z) \Lambda_{(k)} e = 0$ for $z \in [a_{k-1},a_{k}]$
with $k=1,...,N+1$.
We can now define the $N+1$ vectors
$v_{(k)} = \Lambda_{(k)} e $, $k=1,...,N+1$.
Clearly, then these $N+1$ vectors $v_{(k)}$ obey \eqref{e:defv}.
Thus, we have shown that we can always find
$N+1$ vectors $v_{(k)}$ obeying \eqref{e:defv}.

To see that each of the $N+1$ vectors $v_{(k)}$ are unique, seen as elements
of $\R P^{D-3}$, it is enough to notice that we know from
Section \ref{s:rods} that $\dim ( \ker ( G(0,z) )) = 1$ for
$z \in \, ] a_{k-1},a_{k} [$.

%%%%%%%%%%%%%%%%%%%%%%%%%%%%%%%%
\subsubsection*{Discussion of existence and uniqueness of solutions}

We discuss here whether a solution is uniquely given by
its rod-structure, and whether there exists a solution
for any given rod-structure.
We consider here only solutions of Euclidean signature,
but one can easily extend the considerations to solutions of
Lorentzian signature.

If we consider
the special case of the generalized Weyl solutions of \cite{Emparan:2001wk},
corresponding to $G(r,z)$
being a diagonal matrix, we clearly have the directions of the rods
can be chosen to have the form
$v_{(k)} = (0,...,0,\pm 1,0,...,0)$.
It is then known from the analysis of \cite{Emparan:2001wk} that we can
specify a solution completely by the $N$
parameters $a_1 < ... < a_N$ and $N+1$ vectors $v_{(k)}$,
i.e. a solution is completely specified by its rod-structure.

We now speculate that this statement can be generalized,
i.e. that also in the more general class of solution considered
here, a solution is specified uniquely by its rod-structure.
Thus, we claim in detail that:
A solution with $N+1$ rods is completely determined by specifying
the parameters $a_1 < ... < a_N$ and directions $v_{(k)}$, $k=1,...,N+1$.
I.e. it is not possible to find two physically different
solutions with $N+1$ rods that have the same $N$ parameters $a_k$
and $N+1$ directions $v_{(k)}$.
Intuitively, this statement seems valid since one would expect
that the system of equations \eqref{Gijeqs} determine $G(r,z)$
once we have determined the sources for $G(r,z)$ at $r=0$. And,
the values $a_k$ and directions $v_{(k)}$ seems
precisely to specify that.

Note that if this statement is true, it is moreover true that it is
not possible to find two physically different solutions with
$N+1$ rods that have the same $N$ parameters $a_k$, up to a global translation
of all $N$ parameters,
and the same $N+1$ directions $v_{(k)}$, up to a global rotation of
all $N+1$ directions.

One can also turn things
around and ask whether there exists a solution with $N+1$ rods given
the $N$ parameters $a_1 < ... < a_N$ and $N+1$
directions $v_{(k)}$, $k=1,...,N+1$
(not imposing the solution to be regular).
This would be interesting to examine further.
However, there is an obvious restriction on the directions of the
first and last rod $[-\infty,a_1]$ and $[a_N,\infty]$.
For a given asymptotic space, which the solution is required
to asymptote to for
$\sqrt{r^2+z^2} \rightarrow \infty$ with $z/\sqrt{r^2+z^2}$ fixed
(see Section \ref{s:asymp}), the directions of these
two rods should be correlated, and can therefore not be chosen
independently.

%%%%%%%%%%%%%%%%%%%%%%%%%%%%%%%%%%%%%%%%%%%%%%%%%%%%%%%%%%%%%%
\subsection{Analysis of the rod-structure}
\label{s:sumrod}

In this section we summarize how to analyze the rod-structure,
and add some useful nomenclature and general comments.
We consider here solutions $G_{ij} (r,z)$ of \eqref{Gijeqs}
with $\det G = r^2$.

In Section \ref{s:rods} we learned that in order to avoid curvature
singularities it is a necessary condition on a solution
that the eigenspace for the eigenvalue zero of the matrix $G(0,z)$
for a given $z$ should be one-dimensional, except for isolated
values of $z$.
We therefore restrict ourselves to solutions where this applies.
Naming the isolated $z$-values as $a_1,...,a_N$, we see that
the $z$-axis splits up into the $N+1$ intervals $[-\infty,a_1]$,
$[a_1,a_2]$,...,$[a_{N-1},a_N]$, $[a_N,\infty]$.
The first task in understanding the rod-structure
of a solution is thus to find these intervals, called rods.

Consider now a specific rod $[z_1,z_2]$.
From Theorem \ref{theorod} (see also Section \ref{s:genrod})
we know that we can find a vector
\begin{equation}
v = v^i \frac{\partial}{\partial x^i}  \ ,
\end{equation}
so that
\begin{equation}
\label{Gveq}
\sum_{j=1}^{D-2} G_{ij} (0,z) \, v^j = 0 \ ,
\end{equation}
for $i=1,...,D-2$ and $z \in [z_1,z_2]$.
This vector $v$ is called the direction of the rod $[z_1,z_2]$.
Then, if $G_{ij} v^i v^j / r^2$ is negative (positive)
for $r \rightarrow 0$
we say the rod $[z_1,z_2]$ is {\sl time-like} ({\sl space-like}).

Consider now a space-like rod $[z_1,z_2]$.
For $r \rightarrow 0$ with $z \in \, ] z_1,z_2 [$ we have a potential
conical singularity.
Let $\eta$ be a coordinate,
made as a linear combination of $x^i$, $i=1,...,D-2$, with
\begin{equation}
\frac{\partial}{\partial \eta} = v = v^i \frac{\partial}{\partial x^i} \ .
\end{equation}
Then in order to cure the conical singularity at the rod,
the coordinate $\eta$ should have period
\begin{equation}
\label{deta}
\Delta \eta = 2\pi \lim_{r \rightarrow 0}
\sqrt{ \frac{r^2 e^{2\nu} }{G_{ij} v^i v^j} } \ ,
\end{equation}
with $z \in [z_1,z_2]$.
This is seen from the analysis of Section \ref{s:rods}.
We see from this that a space-like rod corresponds to a compact
direction.
For a time-like rod, one can similarly find an associated
temperature, by doing a Wick rotation. Therefore, a time-like
rod corresponds to a horizon (see Section \ref{s:rods}).

We introduce here some additional nomenclature for rods. Consider
a rod $[z_1,z_2]$. If this is a finite interval we call
$[z_1,z_2]$ a finite rod. If either $z_1 = -\infty$ or $z_2 =
\infty$ but not both of them, we call $[z_1,z_2]$ a semi-infinite
rod. $[-\infty,\infty]$ is instead called the infinite rod.

As discussed in \cite{Emparan:2001wk}, a finite time-like rod
corresponds to an event horizon, at least if there are no
semi-infinite time-like rods for the solution. Similarly, a finite
space-like rod corresponds to a Kaluza-Klein direction if there
are no semi-infinite space-like rods in that direction. Moreover,
a (semi-)infinite space-like rod corresponds to an axis of
rotation, with the associated coordinate being the rotation angle,
while a semi-infinite time-like rod corresponds to an acceleration
horizon.

%%%%%%%%%%%%%%%%%%%%%%%%%%%%%%%%%%%%%%%%%%%%%%%%%%%%%%%%%%%%%%
\section{Asymptotically flat space-times}
\label{s:asymp}

In this section we consider asymptotically flat space-times.
We consider the four- and five-dimensional Minkowski-spaces
$\CM^4$ and $\CM^5$, and we consider the asymptotic
behavior of solutions that asymptote to $\CM^4$ and $\CM^5$.

The Minkowski-spaces $\CM^4$ and $\CM^5$ are special in that
they are the only Minkowski-spaces that one can describe
using the ansatz \eqref{themet}-\eqref{rdetG}.
This is easily seen by counting the number of Killing vector
fields.
An obvious generalization of the considerations of this section
would be to consider the Kaluza-Klein space-times $\CM^4 \times S^1$
and $\CM^5 \times S^1$, or other space-times with even more compact
directions, i.e. $\CM^4 \times T^p$ or $\CM^5 \times T^p$.
We leave this for the future.%
\footnote{Strictly speaking, one can consider higher-dimensional
Minkowski-spaces $\CM^D$ with $D \geq 6$, for example by making the
split up $\CM^D = \CM^4 \times \R^{D-4}$, with the $\R^{D-4}$
part spanned by the Killing vector fields. However, one can not use that
to write any non-trivial solutions which asymptotes to $\CM^D$, since
any solution would be independent of the $\R^{D-4}$ part.
Instead, one should consider $\CM^4 \times T^{D-4}$ or $\CM^5 \times
T^{D-5}$.}

In the following we put Newtons constant $G_{\rm N}=1$.
To reinstate $G_{\rm N}$ one should substitute $M \rightarrow G_{\rm N} M$
and $J \rightarrow G_{\rm N} J$.

%%%%%%%%%%%%%%%%%%%%%%%%%%%%%%%%%%%%%%%%%%%%%%%%
\subsection{Perturbation of diagonal metric}
\label{s:pertmet}

Before describing asymptotically flat spaces, we first develop a
tool that will prove useful. We consider in this section
a perturbation $\delta G(r,z)$ of a solution
$G(r,z)$ of Eqs.~\eqref{Gijeqs}, with $G(r,z)$ being diagonal, such that
$G (r,z) + \delta G (r,z)$
also is a solution of Eqs.~\eqref{Gijeqs}.
This will be useful below
since asymptotic behavior of a solution
typically involves the solution asymptoting
towards a diagonal metric like for example the metric of Minkowski-space.
The results here can also be used in a broader context
to find corrections to solutions.

Now, Eqs.~\eqref{Gijeqs} for the perturbation
$\delta G(r,z)$ becomes
\begin{equation}
\label{deltaG}
\grad^2 \delta G_{ij} = \left( \frac{\grad G_{ii}}{G_{ii}}
+ \frac{\grad G_{jj}}{G_{jj}} \right) \cdot \grad \delta  G_{ij}
- \frac{\grad G_{ii}}{G_{ii}} \cdot \frac{\grad G_{jj}}{G_{jj}} \delta  G_{ij}
\ .
\end{equation}
We see here that the equations for $\delta G_{ij}$ are completely
decoupled. I.e. we can solve for each component of $\delta G(r,z)$
separately. The only constraint is that $|\det (G+\delta G)| = r^2$.
Using that $|\det G | = r^2$ this constraint can be written as
$\mbox{tr} ( G^{-1} \delta G ) = 0 $
which we again can write as
\begin{equation}
\label{condelG}
\sum_{i=1}^{D-2} \frac{\delta G_{ii}}{G_{ii}} = 0 \ .
\end{equation}
We see thus that only the diagonal components of $\delta G_{ii}$
are subject to a constraint.
We note that for the diagonal components of $\delta G (r,z)$ the
equations \eqref{deltaG} can be written
\begin{equation}
\label{pertdiag}
\grad^2 \left( \frac{ \delta G_{ii}}{G_{ii}} \right) = 0 \ ,
\end{equation}
for $i=1,...,D-2$.

%%%%%%%%%%%%%%%%%%%%%%%%%%%%%%%%%%%%%%%%%%%%%%%%
\subsection{Four-dimensional asymptotic Minkowski-space}
\label{s:D4mink}

We consider in this section the four-dimensional Minkowski-space
$\CM^4$ and the asymptotic structure of solutions asymptoting
to $\CM^4$.

We first describe $D=4$ Minkowski-space $\CM^4$.
In terms of $G(r,z)$ we have that $\CM^4$ is given by
\begin{equation}
\label{4Dmink}
G_{11} = -1 \spa G_{22} = r^2 \ ,
\end{equation}
Thus, we have an infinite space-like rod $[-\infty,\infty]$.
In accordance with \eqref{nueqs} we choose $e^{2\nu} = 1$.
Demanding regularity of the solution near $r=0$
we get using \eqref{deta}
that $x^2=\phi$ should have period $2\pi$.
Making the coordinate transformation
\begin{equation}
r = \rho \sin \theta
\spa
z = \rho \cos \theta \ ,
\end{equation}
we get the metric in spherical coordinates
\begin{equation}
ds^2 = - dt^2 + \rho^2 \sin^2 \theta d\phi^2
+ d\rho^2 + \rho^2 d\theta^2 \ ,
\end{equation}
where we put $x^1 = t$ and $x^2 = \phi$.

If we consider a $D=4$ asymptotically Minkowski-space solution
we have for $\rho \rightarrow \infty$ the corrections
to the metric
\begin{equation}
\label{gtttphi}
g_{tt} = - 1 + \frac{2 M}{\rho} + \CO (\rho^{-2})
\spa
g_{t\phi} = - 2 J \frac{\sin^2 \theta}{\rho}
\left( 1+ \CO (\rho^{-1})\right) \ ,
\end{equation}
with $g_{\phi \phi} = \rho^2 \sin^2 \theta ( 1 + \CO(\rho^{-1}) )$.

In the $(r,z)$ canonical coordinates the asymptotic
region corresponds to $\sqrt{r^2+z^2} \rightarrow \infty$ with
$z/\sqrt{r^2+z^2}$ finite.
In the canonical coordinates we have therefore from
\eqref{gtttphi} the asymptotic behavior
\begin{equation}
\label{D4asymp}
\begin{array}{c} \ds
G_{11} = - 1 + \frac{2 M }{\sqrt{r^2+z^2}}
+\CO( (r^2+z^2)^{-1} )
\spa
G_{12} = - \frac{2 J r^2}{(r^2+z^2)^{\frac{3}{2}}}
 +\CO( (r^2+z^2)^{-1} ) \ ,
\\[4mm] \ds
G_{22} = r^2 \left[ 1 + \frac{2 M}{\sqrt{r^2+z^2}}
+\CO( (r^2+z^2)^{-1} ) \right] \ ,
\end{array}
\end{equation}
for $\sqrt{r^2+z^2} \rightarrow \infty$
with $z/\sqrt{r^2+z^2}$ finite.
For $G_{22}$ we used \eqref{condelG} and \eqref{pertdiag}
of Section \ref{s:pertmet}.
We see from \eqref{D4asymp} that the leading asymptotic behavior of $G(r,z)$
is determined completely from $M$ and $J$.
For $e^{2\nu}$, the asymptotic behavior is simply that
$e^{2\nu} \simeq 1$ for $\sqrt{r^2+z^2} \rightarrow \infty$
with $z/\sqrt{r^2+z^2}$ finite.

%%%%%%%%%%%%%%%%%%%%%%%%
\subsubsection*{The period of $\phi$}

In the above we considered the period of $x^2=\phi$ to be $2\pi$.
We can also consider the more general case
where $x^2=\phi$ has period $\Delta \phi = 2\pi \varepsilon$.
Then the asymptotic behavior of a solution is
\begin{equation}
\label{4Dgenper}
G_{11} \simeq -1 + \frac{2M}{\varepsilon} \frac{1}{\sqrt{r^2+z^2}} \spa
G_{12} \simeq - \frac{2J}{\varepsilon^2} \frac{r^2}{(r^2+z^2)^{\frac{3}{2}}}
\spa
e^{2\nu} \simeq \varepsilon^2 \ ,
\end{equation}
for $\sqrt{r^2+z^2} \rightarrow \infty$
with $z/\sqrt{r^2+z^2}$ finite, where we used here a less precise
notation than above for the sake of brevity.

%%%%%%%%%%%%%%%%%%%%%%%%%%%%%%%%%%%%%%%%%%%%%%%%%%%%%%%
\subsection{Five-dimensional asymptotic Minkowski-space}
\label{s:D5mink}

We consider in this section the five-dimensional Minkowski-space
$\CM^5$ and the asymptotic structure of solutions asymptoting
to $\CM^5$.

We first describe $D=5$ Minkowski-space $\CM^5$.
In terms of $G(r,z)$ we have that $\CM^5$ is described by
\begin{equation}
\label{D5mink}
G_{11} = -1
\spa
G_{22} = \sqrt{r^2+z^2} - z
\spa
G_{33} = \sqrt{r^2+z^2} + z \ .
\end{equation}
This corresponds to two semi-infinite rods $[-\infty,0]$ and
$[0,\infty]$.
In accordance with Eqs.~\eqref{nueqs} we choose
\begin{equation}
e^{2\nu} = \frac{1}{2\sqrt{r^2+z^2}} \ .
\end{equation}
Demanding regularity of the solution near $r=0$
we get using \eqref{deta}
that both $x^2$ and $x^3$ are periodic with period $2\pi$.
Making the coordinate transformation
\begin{equation}
r = \frac{1}{2} \rho^2 \sin 2 \theta
\spa
z = \frac{1}{2} \rho^2 \cos 2 \theta \ ,
\end{equation}
we get the metric in spheroidal coordinates
\begin{equation}
ds^2 = - dt^2 + \rho^2 \sin^2 \theta d\phi^2 + \rho^2 \cos^2 \theta d\psi^2
+ d\rho^2 + \rho^2 d\theta^2 \ ,
\end{equation}
where we put $x^1=t$, $x^2=\phi$ and $x^3=\psi$.
We remind the reader that regularity of the solution
requires both $x^2=\phi$ and $x^3=\psi$ to be periodic with period $2\pi$.

If we consider a $D=5$ asymptotically Minkowski-space solution
we have for $\rho \rightarrow \infty$ the corrections
to the metric
\begin{equation}
\begin{array}{c} \ds
g_{tt} = - 1 + \frac{8 M}{3\pi} \frac{1}{\rho^2}
+ \CO (\rho^{-4}) \ ,
\\[4mm] \ds
g_{t\phi} = - \frac{4 J_1}{\pi} \frac{\sin^2 \theta}{\rho^2}
\left( 1 + \CO (\rho^{-2}) \right)
\spa
g_{t\psi} = - \frac{4 J_2}{\pi} \frac{\cos^2 \theta}{\rho^2}
\left( 1 + \CO (\rho^{-2}) \right) \ ,
\end{array}
\end{equation}
with $g_{\phi\phi} = \rho^2 \sin^2 \theta ( 1 + \CO ( \rho^{-2} ) )$ and
$g_{\psi\psi} = \rho^2 \cos^2 \theta ( 1 + \CO ( \rho^{-2} ) )$.
Using this together with \eqref{deltaG} and \eqref{condelG}, we get
the asymptotics in the $(r,z)$ canonical coordinates
\begin{eqnarray}
\label{5DGasy}
& \ds
G_{11} = - 1 + \frac{4 M}{3\pi} \frac{1}{\sqrt{r^2+z^2}}
+\CO( (r^2+z^2)^{-1} ) \spa
G_{23} = \zeta \frac{r^2}{(r^2+z^2)^{\frac{3}{2}}}
+\CO( (r^2+z^2)^{-1} ) \ , &
\nn \\ & \ds
G_{12} = - \frac{J_1}{\pi} \frac{\sqrt{r^2+z^2}-z}{r^2+z^2}
+\CO( (r^2+z^2)^{-1} )
\, , \
G_{13} = - \frac{J_2}{\pi} \frac{\sqrt{r^2+z^2}+z}{r^2+z^2}
+\CO( (r^2+z^2)^{-1} ) \, , &
\nn \\ & \ds
G_{22} = \left(\sqrt{r^2+z^2}-z \right) \left[ 1 + \frac{2}{3\pi} \frac{M+\eta}{\sqrt{r^2+z^2}} +\CO( (r^2+z^2)^{-1} )
\right] \ , &
\nn \\ & \ds
G_{33} = \left(\sqrt{r^2+z^2}+z \right) \left[ 1 + \frac{2}{3\pi} \frac{M-\eta}{\sqrt{r^2+z^2}} +\CO( (r^2+z^2)^{-1} )
\right] \ , &
\end{eqnarray}
for $\sqrt{r^2+z^2} \rightarrow \infty$
with $z/\sqrt{r^2+z^2}$ finite, where $\zeta$ and $\eta$
are constants.
Note that $\eta$ changes under the transformation
$z \rightarrow z + \mbox{constant}$ and is thus not
a gauge-invariant parameter, unlike $\zeta$.
Finally, we remark that the asymptotics of $e^{2\nu}$ is
\begin{equation}
e^{2\nu} \simeq \frac{1}{2\sqrt{r^2+z^2}} \ ,
\end{equation}
for $\sqrt{r^2+z^2} \rightarrow \infty$ with $z/\sqrt{r^2+z^2}$ finite.

%%%%%%%%%%%%%%%%%%%%%%%
\subsubsection*{The periods of $\phi$ and $\psi$}

The periods of $x^2=\phi$ and $x^3=\psi$ were chosen to be
$2\pi$ in the above.
We consider here the more general case where
the period $\Delta \phi$ of $x^2=\phi$ and the period $\Delta \psi$
of $x^3=\psi$ are given by $\Delta \phi = \Delta \psi = 2\pi \varepsilon$.
Then the asymptotics of $G_{ij}(r,z)$ and $e^{2\nu}$ takes the form
\begin{equation}
\label{altasympt}
\begin{array}{c} \ds
G_{11} \simeq - 1 + \frac{4 M}{3\pi \varepsilon^2} \frac{1}{\sqrt{r^2+z^2}}
\spa
G_{23} \simeq \frac{\zeta}{\varepsilon^4} \frac{r^2}{(r^2+z^2)^{\frac{3}{2}}}
\spa
e^{2\nu} \simeq  \frac{\varepsilon^2}{2\sqrt{r^2+z^2}} \ ,
\\[4mm] \ds
G_{12} \simeq - \frac{J_1}{\pi \varepsilon^3} \frac{\sqrt{r^2+z^2}-z}{r^2+z^2}
\spa
G_{22} \simeq \left(\sqrt{r^2+z^2}-z \right) \left[ 1 +
\frac{2}{3\pi\varepsilon^2} \frac{M+\eta}{\sqrt{r^2+z^2}} \right] \ ,
\\[4mm] \ds
G_{13} \simeq - \frac{J_2}{\pi \varepsilon^3} \frac{\sqrt{r^2+z^2}+z}{r^2+z^2}
\spa
G_{33} \simeq \left(\sqrt{r^2+z^2}+z \right) \left[ 1 +
\frac{2}{3\pi \varepsilon^2} \frac{M-\eta}{\sqrt{r^2+z^2}} \right] \ ,
\end{array}
\end{equation}
for $\sqrt{r^2+z^2} \rightarrow \infty$
with $z/\sqrt{r^2+z^2}$ finite, where we used here a less precise
notation than above for the sake of brevity.

%%%%%%%%%%%%%%%%%%%%%%%%%%%%%%%%%%%%%%%%%%%%%%%%%%%%%%%%%%%%%%
\section{Rotating black hole solutions}
\label{s:rotBH}

In this section we consider rotating black hole solutions in
four and five dimensions and
describe them using the canonical form of the metric
\eqref{themet}-\eqref{rdetG}.

%%%%%%%%%%%%%%%%%%%%%%%%%%%%%%%%%%%%%%%%
\subsection{Kerr solution}
\label{s:Kerrsol}

We first consider the four-dimensional Kerr solution \cite{Kerr:1963ud}
which corresponds to a rotating black hole. The topology of the event
horizon is that of a two-sphere $S^2$.
It is already known how to write the Kerr solution in
the canonical form \eqref{themet}-\eqref{rdetG}
(see for example \cite{Stephani:2003,Chandrasekhar:1992,Heusler:1996}),
but we review
this here for completeness, and since it illustrates
the methods developed in Section \ref{s:nearrod}.

The Kerr metric in Boyer-Linquist coordinates is
\begin{equation}
\label{kerrBL}
\begin{array}{rcl}
ds^2 &=& \ds
- \frac{\Delta - a^2 \sin^2 \theta }{\Sigma} dt^2
- 2 a \sin^2 \theta \frac{\rho^2 + a^2 - \Delta}{\Sigma} dt d\phi
\\[3mm] && \ds
+ \frac{(\rho^2+a^2)^2 - \Delta a^2 \sin^2 \theta}{\Sigma}
\sin^2 \theta d\phi^2
+ \frac{\Sigma}{\Delta} d \rho^2 + \Sigma d\theta^2 \ ,
\end{array}
\end{equation}
with
\begin{equation}
\label{delsig}
\Delta = \rho^2 - 2M \rho + a^2
\spa
\Sigma = \rho^2 + a^2 \cos^2 \theta \ .
\end{equation}
The coordinates for the
two Killing directions are $x^1 = t$ and $x^2 = \phi$.
From $\det (G) = - \Delta \sin^2 \theta$ we get the $r$-coordinate,
and it is a straightforward
exercise to find a $z$-coordinate so that the metric
fits into the ansatz \eqref{themet}. We find
\begin{equation}
\label{kerrrz}
r = \sqrt{\Delta} \sin \theta \spa
z = (\rho - M) \cos \theta \ .
\end{equation}
Using this, we can in principle write the Kerr metric
in the canonical form \eqref{themet}-\eqref{rdetG}.
However, it is useful to instead first
write the Kerr-metric in the prolate spherical coordinates
(see Appendix \ref{a:prolate}).
From the definition \eqref{defprol}  of
the prolate spherical coordinates $(x,y)$ we see that
\begin{equation}
\alpha^2 (x^2-1)(1-y^2) = \Delta \sin^2 \theta
\spa
\alpha \, x \, y = (\rho-M)\cos \theta \ ,
\end{equation}
Using the ansatz $x=x(\rho)$ and $y=y(\theta)$ we get
\begin{equation}
x = \frac{\rho-M}{\sqrt{M^2-a^2}} \spa
y = \cos \theta \spa
\alpha = \sqrt{M^2 - a^2} \ .
\end{equation}
We compute
\begin{equation}
\label{Kerr1112}
\begin{array}{c} \ds
G_{11} = - \frac{x^2 \cos^2 \lambda + y^2 \sin^2 \lambda -1}{\left( 1 + x \cos \lambda \right)^2 +  y^2 \sin^2 \lambda}
\spa
G_{12} = - 2 a  \frac{(1-y^2)(1 + x \cos \lambda ) }{\left( 1 + x \cos \lambda \right)^2 +  y^2 \sin^2 \lambda } \ ,
\\[4mm] \ds
e^{2\nu } = \frac{\left( 1 + x \cos \lambda \right)^2 +  y^2 \sin^2 \lambda}{(x^2-y^2)\cos^2 \lambda} \ ,
\end{array}
\end{equation}
where we defined
\begin{equation}
\sin \lambda = \frac{a}{M} \ .
\end{equation}
The $G_{22}$ component can be found from
\begin{equation}
\label{Kerr22}
G_{22} = \frac{G_{12}^2 - \alpha^2 (x^2-1)(1-y^2)}{G_{11}} \ .
\end{equation}
We obtained now $G_{ij}$ and $e^{2\nu}$ as functions of $x$ and $y$.
From this it is straightforward
to use Eq.~\eqref{xyfrz} to get $G_{ij}$ and $e^{2\nu}$
as functions of $r$ and $z$.

%%%%%%%%%%%%%%%%%%%%%%%%
\subsubsection*{Asymptotic region}

Using \eqref{xyrzasymp} we find that in the asymptotic region
$\sqrt{r^2+z^2} \rightarrow \infty$ with $z/\sqrt{r^2+z^2}$ finite,
we have
\begin{equation}
G_{11} = - 1 + \frac{2M}{\sqrt{r^2+z^2}} + \CO ( (r^2+z^2)^{-1} )
\spa
G_{12} = - 2Ma \frac{ r^2}{(r^2+z^2)^{3/2}} + \CO ( (r^2+z^2)^{-1} ) \ .
\end{equation}
From \eqref{D4asymp} we see that this means that
$M$ is the mass, which justifies our use of this symbol
in the solution,
and that the angular momentum is $J = M a$.

Note that $e^{2\nu} \simeq 1$ in the asymptotic region.
From Section \ref{s:D4mink} we know that this
means that $\Delta \phi = 2\pi$,
i.e. that $\phi$ is required to have period $2\pi$. This
can also be found directly from the solution near $r=0$
using the analysis of Section \ref{s:rods}.

%%%%%%%%%%%%%%%%%%%%%%%%%
\subsubsection*{Rod-structure}

We now analyse the rod-structure of the Kerr solution according
to the methods of Section \ref{s:nearrod}. We have:
\begin{itemize}
\item The two semi-infinite space-like rods $[-\infty,-\alpha]$
and $[\alpha,\infty]$.
For $z \in [-\infty,-\alpha]$ and $r=0$ we see from \eqref{xyfrz}
that $x = - z/\alpha$ and $y=-1$. Similarly,
for $z \in [\alpha,\infty]$ and $r=0$ we have
that $x = z/\alpha$ and $y=1$.
Considering Eqs.~\eqref{Kerr1112} and \eqref{Kerr22} we see that
for both intervals
$G_{12} = G_{22} = 0$ while $G_{11} \neq 0$. By Eq.~\eqref{Gveq}
this means
that the two rods both are in the direction
$v = (0,1)$, i.e. that the rods
are in the $\partial/\partial x^2$
direction and therefore space-like.
\item The finite time-like rod $[-\alpha,\alpha]$.
For $z \in [-\alpha,\alpha]$ we see from \eqref{xyfrz} that
$x = 1$ and $y = z/\alpha$.
Considering Eqs.~\eqref{Kerr1112} and \eqref{Kerr22} we see that
$\sum_{j=1}^2 G_{ij} v^j = 0$ for $z \in [-\alpha,\alpha]$ with
\begin{equation}
\label{vkerr}
v = ( 1  , \Omega )
\spa
\Omega = \frac{\sin \lambda}{2M ( 1 + \cos \lambda )} \ .
\end{equation}
This means that we have a rod $[-\alpha,\alpha]$ along
the direction \eqref{vkerr}.
Since $G_{ij} v^i v^j/r^2$ is negative for $r \rightarrow 0$ the
rod is time-like.
Note that $\Omega$ in \eqref{vkerr} is the
{\sl angular velocity} of the event horizon.%
\footnote{Note that $v$ in \eqref{vkerr} precisely is the null Killing vector
for the Killing horizon (the event horizon),
since $v = \sum_{i=1}^2 v^i V_{(i)}$,
and since $v^2 = G_{ij}v^i v^j = 0$ for $r=0$ and $z \in [-\alpha,\alpha]$.
In other words, for a Killing horizon the null Killing vector is the same as
the direction of the time-like rod.}
One finds easily that this rod corresponds to an event horizon
of topology $S^2$. This is a consequence of the fact
that the rods on each side of the $[-\alpha,\alpha]$ rod are in
the same space-like direction, i.e. the $\partial/\partial x^2$ direction.
\end{itemize}

For the time-like rod $[-\alpha,\alpha]$ we see from
\eqref{vkerr} that if we change coordinates as
$\tilde{x}^1 = x^1$ and
$\tilde{x}^2 = x^2 - \Omega x^1 $,
then in these coordinates the $[-\alpha,\alpha]$
rod is along the $\partial/ \partial \tilde{x}^1$ direction.
This means that $\tilde{x}^1$ and $\tilde{x}^2$
are two of the coordinates of the comoving coordinates
for the Kerr solution since
the comoving coordinates precisely gives a diagonal metric at the horizon.
In other words, finding the direction of the $[-\alpha,\alpha]$
rod precisely corresponds to finding
the comoving coordinates near the horizon.

Finally, in the accordance with the ideas of Section \ref{s:genrod},
we note that we can make an alternative parameterization of the Kerr
solution by stating that we have three rods, the two rods
$[-\infty,-\alpha]$ and $[\alpha,\infty]$ in the $x^2$ direction
and the rod $[-\alpha,\alpha]$ in the $(1, \Omega)$ direction.
Then the whole Kerr solution can be parameterized uniquely by
the two parameters $\alpha$ and $\Omega$.

%%%%%%%%%%%%%%%%%%%%%%%%%%%%%%%%%%%%%%%%%%%%%%%%%%
\subsection{Five-dimensional Myers-Perry solution}
\label{s:MPsol}

The five-dimensional Myers-Perry solution \cite{Myers:1986un} corresponds to a
five-dimensional spinning black hole.%
\footnote{The five-dimensional Myers-Perry black hole
generalizes the static Schwarzschild-Tangherlini black hole
\cite{Tangherlini:1963}.}
This is an asymptotically flat
stationary solution of the vacuum Einstein equations
with an event horizon
that has the topology of a three-sphere $S^3$.

The metric of the five-dimensional Myers-Perry black hole is
\begin{equation}
\label{MPmet}
\begin{array}{rcl}
ds^2 &=& \ds - dt^2
+ \frac{\rho_0^2}{\Sigma} \left[ dt - a_1 \sin^2 \theta d\phi
- a_2 \cos^2 \theta d\psi \right]^2
\\ && \ds
+ (\rho^2 + a_1^2) \sin^2 \theta d\phi^2
+ (\rho^2 + a_2^2) \cos^2 \theta d\psi^2
+ \frac{\Sigma}{\Delta} d\rho^2 + \Sigma d\theta^2 \ ,
\end{array}
\end{equation}
where
\begin{equation}
\Delta = \rho^2 \left( 1 + \frac{a_1^2}{\rho^2} \right)
\left( 1 + \frac{a_2^2}{\rho^2} \right) - \rho_0^2
\spa
\Sigma = \rho^2 + a_1^2 \cos^2 \theta + a_2^2 \sin^2 \theta \ .
\end{equation}
The coordinates for the
three Killing directions are $x^1 = t$, $x^2 = \phi$ and $x^3 = \psi$.
We now transform this metric to the canonical
form \eqref{themet}-\eqref{rdetG}.
We compute that $\det G = - \frac{1}{4} \rho^2 \Delta \sin^2 2\theta$.
From this we can determine $r$, and $z$ can be found by demanding
the metric to be of the form \eqref{themet}. We get
\begin{equation}
\label{rztwo}
r = \frac{1}{2} \rho \sqrt{\Delta} \sin 2\theta
\spa
z = \frac{1}{2} \rho^2 \left( 1 - \frac{\rho_0^2 - a_1^2 - a_2^2}{2\rho^2} \right)
\cos 2 \theta \ .
\end{equation}
This determines in principle how the Myers-Perry metric \eqref{MPmet}
should transform to the form \eqref{themet}. However, as for the
Kerr metric, it is convinient to express the Myers-Perry metric
in prolate spherical coordinates, defined
in Appendix \ref{a:prolate}, instead.
From the definition of
the prolate spherical coordinates \eqref{defprol} we see that
\begin{equation}
\alpha^2 (x^2-1)(1-y^2) = \frac{1}{4} \rho^2 \Delta \sin^2 2\theta
\spa
\alpha \, x \, y = \frac{1}{2} \rho^2 \left( 1 - \frac{\rho_0^2 - a_1^2 - a_2^2}{2\rho^2} \right)
\cos 2 \theta \ .
\end{equation}
If we try the ansatz $x=x(\rho)$ and $y=y(\theta)$ we get
\begin{equation}
\label{xytwo}
x = \frac{2 \rho^2 + a_1^2 + a_2^2 - \rho_0^2}{\sqrt{ ( \rho_0^2 - a_1^2 - a_2^2 )^2
- 4 a_1^2 a_2^2 }}
\spa
y = \cos 2 \theta
\spa
\alpha = \frac{1}{4} \sqrt{ ( \rho_0^2 - a_1^2 - a_2^2 )^2
- 4 a_1^2 a_2^2 } \ .
\end{equation}
Using this, we can write $G_{ij}$ and $e^{2\nu}$ in terms of the prolate
spherical coordinates. We get
\begin{equation}
\label{MPmetxy}
\begin{array}{c} \ds
G_{11} = - \frac{4\alpha x + ( a_1^2 - a_2^2 ) y - \rho_0^2}{4\alpha x + ( a_1^2 - a_2^2 ) y + \rho_0^2}
\spa
G_{12} = - \frac{a_1 \rho_0^2 (1-y)}{4\alpha x
+ (a_1^2-a_2^2) y + \rho_0^2} \ ,
\\[4mm] \ds
G_{13} = - \frac{a_2 \rho_0^2 (1+y)}{4\alpha x
+ (a_1^2-a_2^2) y + \rho_0^2}
\spa
G_{23} = \frac{1}{2} \frac{a_1 a_2 \rho_0^2 (1-y^2) }{4 \alpha x + (a_1^2 - a_2^2 ) y + \rho_0^2} \ ,
\\[4mm] \ds
G_{22} = \frac{1-y}{4} \left[ 4 \alpha x
+ \rho_0^2 + a_1^2-a_2^2
+ \frac{2 a_1^2 \rho_0^2 (1-y)}{4 \alpha x + (a_1^2 - a_2^2 ) y + \rho_0^2}
\right] \ ,
\\[4mm] \ds
G_{33} = \frac{1+y}{4} \left[ 4 \alpha x
+ \rho_0^2 - a_1^2 + a_2^2
+ \frac{2 a_2^2 \rho_0^2 (1+y)}{4 \alpha x + (a_1^2 - a_2^2 ) y + \rho_0^2}
\right] \ ,
\\[4mm] \ds
e^{2\nu} = \frac{4\alpha x + (a_1^2-a_2^2)y + \rho_0^2}{8\alpha^2(x^2-y^2)}
\ .
\end{array}
\end{equation}
Using Eq.~\eqref{xyfrz} it is
now a straightforward exercise to write the components
$G_{ij}$ and $e^{2\nu}$ as
functions of the canonical $(r,z)$ coordinates.

%%%%%%%%%%%%%%%%%%%%%%%%%%%%%
\subsubsection*{Asymptotic region}

Regarding \eqref{MPmetxy} as functions of the
canonical coordinates $(r,z)$, we find that
$G_{ij} (r,z)$ in the asymptotic region, $\sqrt{r^2+z^2}\rightarrow \infty$
with $z/\sqrt{r^2+z^2}$ finite, behaves as
\begin{eqnarray}
& \ds
G_{11} = -1 + \frac{\rho_0^2}{2 \sqrt{r^2+z^2}} + \CO( (r^2+z^2)^{-1} )
, \
G_{12} = - \frac{a_1\rho_0^2}{4} \frac{\sqrt{r^2+z^2}-z}{r^2+z^2}
+ \CO( (r^2+z^2)^{-1} ) , &
\nn \\
& \ds
G_{13} = - \frac{a_2\rho_0^2}{4} \frac{\sqrt{r^2+z^2}+z}{r^2+z^2}
+ \CO( (r^2+z^2)^{-1} )
\spa
G_{23} = \frac{a_1 a_2 \rho_0^2 r^2}{8(r^2+z^2)^{3/2}}
+ \CO( (r^2+z^2)^{-1} ) \ , &
\nn \\
& \ds
G_{22} = \left( \sqrt{r^2+z^2} - z \right) \left[
1 + \frac{\rho_0^2 + a_1^2 - a_2^2}{4 \sqrt{r^2+z^2}}
+ \CO( (r^2+z^2)^{-1} ) \right] \ , &
\nn \\
& \ds
G_{33} = \left( \sqrt{r^2+z^2} + z \right) \left[
1 + \frac{\rho_0^2 - a_1^2 + a_2^2}{4 \sqrt{r^2+z^2}}
+ \CO( (r^2+z^2)^{-1} ) \right] \ . &
\end{eqnarray}
We can now use \eqref{5DGasy} to read off the
asymptotic quantities. We get
\begin{equation}
\label{MPasy}
M = \frac{3\pi}{8} \rho_0^2 \spa
J_1 = \frac{\pi}{4} a_1 \rho_0^2 \spa
J_2 = \frac{\pi}{4} a_2 \rho_0^2 \spa
\zeta = \frac{1}{8} a_1 a_2 \rho_0^2 \spa
\eta = \frac{3\pi}{8} ( a_1^2 - a_2^2 ) \ .
\end{equation}
Note that one can see from the above results
that $e^{2\nu} \simeq 1/(2\sqrt{r^2+z^2})$.
From Section \ref{s:D5mink} we have that this means $x^2=\phi$
and $x^3=\psi$ are periodic with period $2\pi$.

%%%%%%%%%%%%%%%%%%%%%%%%%%%%%
\subsubsection*{Rod-structure}

We now analyse the rod-structure of the five-dimensional
Myers-Perry solution according
to the methods of Section \ref{s:nearrod}.
We have
\begin{itemize}
\item The semi-infinite space-like rod $[-\infty,-\alpha]$.
For $z \in [-\infty,-\alpha]$ and $r=0$ we see from \eqref{xyfrz}
that $x = - z/\alpha$ and $y=-1$.
From \eqref{MPmetxy} we see then that
$G_{13} = G_{23} = G_{33} = 0$.
By Eq.~\eqref{Gveq} we see that this rod has the direction $v = (0,0,1)$,
i.e. it is in the $\partial / \partial x^3$ direction.
\item The finite time-like rod $[-\alpha,\alpha]$.
For $z \in [-\alpha,\alpha]$ we see from \eqref{xyfrz} that
$x = 1$ and $y = z/\alpha$.
Using \eqref{MPmetxy}, we see that
$\sum_{j=1}^3 G_{ij} v^j = 0$ for $z \in [-\alpha,\alpha]$
with $v$ being the vector
\begin{equation}
\label{MPv}
v = (1 , \Omega_1 , \Omega_2 ) \ ,
\end{equation}
with
\begin{equation}
\label{MPoms}
\Omega_1 = \frac{\rho_0^2 + a_1^2 - a_2^2 - 4\alpha}{2a_1\rho_0^2}
\spa
\Omega_2 = \frac{\rho_0^2 - a_1^2 + a_2^2 - 4\alpha}{2a_2\rho_0^2} \ .
\end{equation}
Therefore, the rod $[-\alpha,\alpha]$
is in the direction $v$ given by \eqref{MPv}.
Note that $\Omega_1$ and $\Omega_2$ are the angular velocities
of the Myers-Perry black hole.
That the rod $[-\alpha,\alpha]$ is time-like can be seen by noting
that $G_{ij} v^i v^j / r^2$ is negative for $r\rightarrow 0$.
One can check that this rod corresponds to an
event horizon with topology $S^3$. This follows
from the fact that the $[-\alpha,\alpha]$ rod has
space-like rods on each side in two different directions.
\item The semi-infinite space-like rod $[\alpha,\infty]$.
For $z \in [\alpha,\infty]$ and $r=0$ we see
that $x = z/\alpha$ and $y=1$.
From \eqref{MPmetxy} we see then that
$G_{12} = G_{22} = G_{23} = 0$.
This means that the rod $[\alpha,\infty]$ is in the direction $v = (0,1,0)$,
i.e. in the $\partial / \partial x^2$ direction.
\end{itemize}

We see that one can use the direction
\eqref{MPv} to transform to coordinates
$(\tilde{x}^1,\tilde{x}^2,\tilde{x}^3)
= (x^1,x^2-\Omega_1 x^1,x^3 - \Omega_2 x^1)$
so that the $[-\alpha,\alpha]$ rod
is along the $\partial / \partial \tilde{x}^1$ direction.
This means that $(\tilde{x}^1,\tilde{x}^2,\tilde{x}^3)$
are comoving coordinates for the event horizon.
Thus, finding the direction of the rod $[-\alpha,\alpha]$
corresponds to finding the comoving coordinates on the horizon
of the Myers-Perry black hole.

We note furthermore that we can make an alternative parameterization
of the five-dimensional Myers-Perry solution.
Clearly, the direction $v=(1,\Omega_1,\Omega_2)$
in \eqref{MPv} is given uniquely by the
two parameters $\Omega_1$ and $\Omega_2$.
Letting then $\alpha$ be the third parameter,
we have that the five-dimensional Myers-Perry solution
is characterized uniquely by the three parameters
$\alpha$, $\Omega_1$ and $\Omega_2$, in
accordance with the ideas of Section \ref{s:genrod}.

%%%%%%%%%%%%%%%%%%%%%%%%%%%%%%%%%%%%%%%%%%%%%%%%%%
\subsection{Myers-Perry solution with one angular momentum}
\label{s:MPone}

In the following we give details on the
five-dimensional Myers-Perry solution with
one angular momentum.
In our conventions, we obtain the Myers-Perry solution with one
angular momentum below from the Myers-Perry solution with
two angular momenta above by setting $a_1 = a$ and $a_2 = 0$.
The metric is
\begin{equation}
\begin{array}{rcl}
ds^2 &=& \ds - dt^2
+ \frac{\rho_0^2}{\Sigma} \left( dt - a \sin^2 \theta d\phi \right)^2
+ (\rho^2 + a^2) \sin^2 \theta d\phi^2
+ \rho^2 \cos^2 \theta d\psi^2
\\[2mm] && \ds
+ \frac{\Sigma}{\Delta} d\rho^2 + \Sigma d\theta^2 \ ,
\end{array}
\end{equation}
where
\begin{equation}
\Delta = \rho^2 - \rho_0^2 + a^2
\spa
\Sigma = \rho^2 + a^2 \cos^2 \theta \ .
\end{equation}
From \eqref{rztwo} we see that the $(r,z)$ coordinates are given by
\begin{equation}
r = \frac{1}{2} \rho \sqrt{\Delta} \sin 2 \theta
\spa
z = \left( \frac{1}{2} \rho^2
- \frac{\rho_0^2 - a^2}{4} \right) \cos 2 \theta \ .
\end{equation}
The prolate spherical coordinates are given by
\begin{equation}
x = \frac{2\rho^2}{\rho_0^2 - a^2} - 1
\spa
y = \cos 2 \theta
\spa
\alpha = \frac{\rho_0^2 - a^2}{4} \ ,
\end{equation}
as one can see from \eqref{xytwo}.
From Eqs.~\eqref{MPmetxy} we obtain that
the metric in prolate spherical coordinates is given by
\begin{equation}
\begin{array}{c} \ds
G_{11} = - \frac{x \cos^2 \lambda + y \sin^2 \lambda - 1}{x \cos^2 \lambda + y \sin^2 \lambda + 1}
\spa
G_{12} = - \frac{2\sqrt{\alpha} \tan \lambda (1-y)}{x \cos^2 \lambda + y \sin^2 \lambda + 1} \ ,
\\[4mm] \ds
G_{22} = \frac{\alpha}{\cos^2 \lambda} (1-y) \left[ x \cos^2 \lambda
+ 1 + \sin^2 \lambda
+ \frac{2 \sin^2 \lambda  (1-y)}{x \cos^2 \lambda + y\sin^2 \lambda  + 1} \right] \ ,
\\[4mm] \ds
G_{33} = \alpha (x+1) (1+y)
\spa
e^{2\nu} = \frac{x \cos^2 \lambda + y \sin^2 \lambda + 1}{2\alpha \cos^2 \lambda (x^2-y^2)} \ ,
\end{array}
\end{equation}
where we have defined
\begin{equation}
\sin \lambda = \frac{a}{\rho_0} \ .
\end{equation}
Using Eq.~\eqref{xyfrz} we get furthermore the
metric written in the canonical form \eqref{themet}-\eqref{rdetG}
as functions of the canonical $(r,z)$ coordinates
\begin{equation}
\label{MPcan}
\begin{array}{c} \ds
G_{11} = - \frac{R_+ + R_- \cos 2\lambda - 2\alpha}{R_+ + R_- \cos 2\lambda + 2\alpha}
\spa
G_{12} = - \frac{2\sqrt{\alpha} \tan \lambda ( 2\alpha - R_+ + R_- )}{R_+ + R_- \cos 2\lambda + 2\alpha} \ ,
\\[4mm] \ds
G_{22} = \frac{2\alpha - R_+ + R_-}{4\alpha}
\left[ R_+ + R_- + 2\alpha \frac{1+\sin^2 \lambda }{\cos^2 \lambda}
+ \frac{4\alpha \tan^2 \lambda (2\alpha - R_+ + R_-)}
{R_+ + R_- \cos 2\lambda + 2\alpha}
\right] \ ,
\\[4mm] \ds
G_{33} = R_+ + z + \alpha
\spa
e^{2\nu} = \frac{R_+ + R_- \cos 2\lambda + 2\alpha}{4 R_+ R_- \cos^2 \lambda}
\ ,
\end{array}
\end{equation}
with
\begin{equation}
R_+ = \sqrt{r^2+(z+\alpha)^2} \spa R_- = \sqrt{r^2+(z-\alpha)^2} \ .
\end{equation}
We see that the whole solution \eqref{MPcan} is
determined by the two parameters $\alpha$ and $\lambda$.

From the analysis of the asymptotic region
of the Myers-Perry solution with two angular momenta, we see that
the asymptotic quantities are now
\begin{equation}
M = \frac{3\pi}{8} \rho_0^2 \spa
J_1 = \frac{\pi}{4} a \rho_0^2 \spa
J_2 = 0 \spa
\zeta = 0 \spa
\eta = \frac{3\pi}{8} a^2 \ .
\end{equation}
We list for completeness here the rod-structure of the solution
\eqref{MPcan}. This can easily be obtained
using the results
of the analysis for the case of two angular momenta.
\begin{itemize}
\item The semi-infinite space-like rod $[-\infty,-\alpha]$.
This rod is in the direction $v = (0,0,1)$,
i.e. in the $\partial / \partial x^3$ direction.
\item The finite time-like rod $[-\alpha,\alpha]$.
This rod is in the direction $v$ given by
\begin{equation}
v = (1,\Omega,0)
\spa
\Omega=\frac{a}{\rho_0^2} = \frac{\sin \lambda \cos \lambda}{2\sqrt{\alpha}}
\ .
\end{equation}
\item The semi-infinite space-like rod $[\alpha,\infty]$.
This rod is in the direction $v = (0,1,0)$,
i.e. in the $\partial / \partial x^2$ direction.
\end{itemize}

%%%%%%%%%%%%%%%%%%%%%%%%%%%%%%%%%%%%%%%%%%%%%%%%%%%%%%%%%%
\section{Black ring solutions}
\label{s:BRsol}

In this section we consider the rotating black ring \cite{Emparan:2001wn}.
The rotating black ring is the first known example of a stationary and
regular asymptotically flat five-dimensional solution with an
event horizon that is not topologically a three-sphere $S^3$.
Instead the horizon is topologically a ring $S^2 \times S^1$.

We first describe in Section \ref{s:genBR}
the general black ring solution
which generically has a conical singularity.
We write its metric in the canonical
form \eqref{themet}-\eqref{rdetG} and discuss
the rod-structure.
We consider then briefly the special case
of the static black ring solution and furthermore how
to obtain the Myers-Perry rotating black hole with one angular
momentum.

In Section \ref{s:rotBR} we present the
regular rotating black ring, write its metric in the canonical
form \eqref{themet}-\eqref{rdetG} and discuss its properties.

%%%%%%%%%%%%%%%%%%%%%%%%%%%%%%%%%%%%%%%%%%%%%%%%%%
\subsection{General black ring solution}
\label{s:genBR}

We begin by reviewing briefly the general black ring metric.
So far, the general black ring metric has been written only
in the so-called C-metric coordinates.
In the C-metric coordinates of \cite{Emparan:2004wy},
the general black ring metric is%
\footnote{We use here the C-metric coordinates
of \cite{Emparan:2004wy}, since they are
particular convinient for our purposes.
There have been given three different, but equivalent, C-metric
coordinates for the black ring in the literature:
i) The original coordinates of \cite{Emparan:2001wn}.
ii) The coordinates described in \cite{Hong:2003gx} where one takes the
solution of \cite{Emparan:2001wn} and rewrite
it so that structure functions are factorizable.
iii) The coordinates of \cite{Emparan:2004wy} where the
structure functions also are factorizable, but where
$\det G$ is simpler.}
\begin{equation}
\label{genBRmet}
ds^2 = - \frac{F(v)}{F(u)} \left( dt
-  C \el \frac{1+v}{F(v)} d\phi \right)^2
+ \frac{2\el^2 F(u)}{(u-v)^2} \left[
- \frac{G(v)}{F(v)} d\phi^2
+ \frac{G(u)}{F(u)} d\psi^2
+ \frac{du^2}{G(u)} - \frac{dv^2}{G(v)}
\right] .
\end{equation}
Here $F(\xi)$ and $G(\xi)$ are the structure functions, which takes
the form
\begin{equation}
\label{strucfcts}
F(\xi) = 1 + b \xi \spa
G(\xi) = (1-\xi^2)(1+ c \xi) \ ,
\end{equation}
where the parameters $b$ and $c$ lie in the ranges
\begin{equation}
0 < c \leq b < 1 \ .
\end{equation}
Furthermore, in \eqref{genBRmet}, the constant $C$ is given in
terms of $b$ and $c$ by
\begin{equation}
\label{conC}
C = \sqrt{2 b (b - c) \frac{1+b}{1-b}} \ .
\end{equation}
The $u$ and $v$ coordinates in \eqref{genBRmet} have the ranges
\begin{equation}
-1 \leq u \leq 1 \spa   v \leq - 1 \ .
\end{equation}
Note that the solution \eqref{genBRmet}
generically has conical singularities at $u=1$, $u=-1$ and $v=-1$
\cite{Emparan:2004wy}.
These will be analyzed below using the methods of Section \ref{s:nearrod}.
We note here that while the potential singularities
at $u=-1$ and $v=-1$ will be cured by choosing the periods
of $x^2=\phi$ and $x^3=\psi$ appropiately, we do not fix the singularity
at $u=1$ before in Section \ref{s:rotBR} where we consider the regular
rotating black ring. Thus, in the following the black ring solution
is generically singular at $u=1$.

%%%%%%%%%%%%%%%%%%%%%%%%%%%%
\subsubsection*{Metric in canonical coordinates}

We now find the metric in the canonical coordinates
\eqref{themet}-\eqref{rdetG}.
In the following we use extensively the results of
Appendix  \ref {s:cmet}. In Appendix \ref {s:cmet}
the general relation between C-metric coordinates $(u,v)$
and the canonical coordinates $(r,z)$ is discussed in detail.
Furthermore, for the specific case relevant here
several useful relations between the C-metric coordinates $(u,v)$ and
the canonical coordinates $(r,z)$ are given.

We take the coordinates for the Killing directions to be
$x^1=t$, $x^2=\phi$ and $x^3=\psi$.
Using the results of Appendix \ref{s:cmet} we see that
the $r$ and $z$ coordinates takes the form
\begin{equation}
r = \frac{2\el^2 \sqrt{- G(u)G(v)}}{(u-v)^2}
\spa
z = \frac{\el^2 (1-u v)(2 + c u+ c v)}{(u-v)^2} \ .
\end{equation}
This is obtained by first computing $\det G$, which gives $r$.
In Appendix \ref {s:cmet} it is then found for this particular
$r$, given by the structure function $G(\xi)$ in \eqref{strucfcts},
that $z$ can be chosen as in \eqref{findz}.

From \eqref{genBRmet} and \eqref{strucfcts}
we get using \eqref{uvfromrz} of Appendix \ref {s:cmet},
giving $u$ and $v$ as functions of $r$ and $z$, that $G_{ij} (r,z)$ is
\begin{equation}
\label{BRgenrz}
\begin{array}{c} \ds
G_{11} = - \frac{(1+b)(1-c)R_1 + (1-b)(1+c)R_2
- 2(b-c)R_3 - 2b (1-c^2)\el^2}{(1+b)(1-c)R_1 + (1-b)(1+c)R_2
- 2(b-c)R_3 + 2b (1-c^2)\el^2} \ ,
\\[5mm] \ds
G_{12} = -
\frac{2C\el (1-c)(R_3-R_1 + (1+c)\el^2)}{(1+b)(1-c)R_1
+ (1-b)(1+c)R_2 - 2(b-c)R_3 + 2b (1-c^2)\el^2} \ ,
\\[5mm] \ds
G_{33} = \frac{(R_1+R_2+2c\el^2)(R_1-R_3+(1+c)\el^2)
(R_2+R_3-(1-c)\el^2)}{2\el^2((1-c)R_1-(1+c)R_2-2cR_3)} \ ,
\end{array}
\end{equation}
where we have defined $R_1$, $R_2$ and $R_3$ by
\begin{equation}
\label{defRs}
R_1 = \sqrt{r^2 + (z+c \el^2)^2} \spa
R_2 = \sqrt{r^2 + (z-c \el^2)^2} \spa
R_3 = \sqrt{r^2 + (z- \el^2)^2} \ ,
\end{equation}
as also defined in \eqref{appdefRszs}.
For simplicity, we do not write $G_{22}$ explicitly here, but
note that it is given implicitly as a function of $(r,z)$ by
\begin{equation}
\label{G22BR}
G_{22} = -\frac{r^2}{G_{11}G_{33}} + \frac{G_{12}^2}{G_{11}} \ .
\end{equation}
Using now furthermore \eqref{nuzeta}, we get
\begin{equation}
\label{e2nuBR}
\begin{array}{rcl}
e^{2\nu} &=& \ds
\left[ (1+b)(1-c) R_1 + (1-b)(1+c)R_2 + 2(c-b)R_3
+ 2b(1-c^2)\el^2 \right]
\\[1mm] &&  \ds \times
\frac{(1-c)R_1 + (1+c)R_2 + 2cR_3}{8(1-c^2)^2 R_1 R_2 R_3} \ .
\end{array}
\end{equation}
This completes the general black ring solution
as written in canonical coordinates
\eqref{themet}-\eqref{rdetG}.
Note that using \eqref{Rszs} in Appendix \ref {s:cmet} it is easy to see
that $G_{33}$ can be written in the alternative form
\begin{equation}
\label{altG33}
G_{33} = \frac{(R_3 + z-\el^2)(R_2 - z+c\el^2)}{R_1 - z-c\el^2} \ .
\end{equation}
%

%%%%%%%%%%%%%%%%%
\subsubsection*{Rod-structure}

We now analyze the rod-structure of the general black ring metric.
This includes an analysis of the possible conical singularities
of the solution. The rod-structure is as follows:
\begin{itemize}
\item The semi-infinite space-like rod $[-\infty,-c\el^2]$.
For $r=0$ and $z\in [-\infty,-c\el^2]$ we have
that $R_1 - R_3 + (1+c)\el^2 = 0$ which using
\eqref{BRgenrz} is seen to give that $G_{33} = 0$.
This means we have a rod $[-\infty,-c\el^2]$ in the
direction $v=(0,0,1)$, i.e. in the $\partial / \partial x^3$ direction.
Using \eqref{deta} we see furthermore
that $x^3=\psi$ needs to have period
\begin{equation}
\label{psiper}
\Delta \psi = 2\pi \frac{\sqrt{1-b}}{1-c} \ ,
\end{equation}
to avoid a conical singularity for $r=0$ and $z\in [-\infty,-c\el^2]$.
Since $u=-1$ is equivalent to $ R_1 - R_3 + (1+c)\el^2 = 0$
we see that this conical singularity corresponds to the one at
$u=-1$ mentioned above.
\item The finite time-like rod $[-c\el^2,c\el^2]$.
For $r=0$ and $z\in [-c\el^2,c\el^2]$ we see that
$R_1 + R_2 - 2 c \el^2 = 0$.
One can then check that $\sum_{j=1}^3 G_{ij} v^j = 0$ for
$r=0$ and $z\in [-c\el^2,c\el^2]$ with $v$ being the vector
\begin{equation}
\label{vgenBR}
v = (1,\Omega,0)
\spa
\Omega = \frac{b-c}{(1-c)C\kappa} \ .
\end{equation}
From this we see that we have a rod $[-c\el^2,c\el^2]$ along
the direction $v$ given in \eqref{vgenBR}.
The rod $[-c\el^2,c\el^2]$ is time-like since
$G_{ij} v^i v^j / r^2$ is negative for $r \rightarrow 0$.
Note that $\Omega$ in \eqref{vgenBR}
is the angular velocity of the general black ring
solution.
One can check that this rod corresponds to an event horizon
with topology $S^2 \times S^1$. This follows
from the fact that the $[-c\el^2,c\el^2]$ rod has
rods in the $\partial/\partial x^3$ direction on each side, so that
the $z$ and $x^3$ coordinates parameterize the $S^2$ while
the $x^2$ coordinate parameterize the $S^1$.
\item The finite space-like rod $[c\el^2,\el^2]$.
For $r=0$ and $z\in [c\el^2,\el^2]$ we have
that $R_2 + R_3 - (1-c)\el^2=0$.
Using \eqref{BRgenrz} we see that this gives that $G_{33}=0$.
This means we have a rod $[c\el^2,\el^2]$
in the direction $v=(0,0,1)$, i.e. in the $\partial / \partial x^3$
direction.
Using \eqref{deta} we see furthermore
that $x^3=\psi$ needs to have period
\begin{equation}
\label{extrapsi}
\Delta \psi = 2\pi \frac{\sqrt{1+b}}{1+c} \ ,
\end{equation}
to avoid a conical singularity for $r=0$ and $z\in [c\el^2,\el^2]$.
However, since we have already fixed the period of $x^3=\psi$ by
\eqref{psiper}, curing the conical singularity associated with
the rod $[c\el^2,\el^2]$ requires putting $b = 2c/(1+c^2)$.
We do not fix $b$ in terms of $c$ here,
thus we consider here solutions that can
have conical singularities for $r=0$ and $z\in [c\el^2,\el^2]$.
In Section \ref{s:rotBR} we consider the subset of solutions
for which we do not have any conical singularities.
Finally, note that since $u=1$ is equivalent to $R_2 + R_3 - (1-c)\el^2=0$
we see that this conical singularity corresponds to the one at
$u=1$ mentioned above.
\item The semi-infinite space-like rod $[\el^2,\infty]$.
For $r=0$ and $z\in [\el^2,\infty]$ we have
that $R_1 - R_3 - (1+c)\el^2 = 0$.
Using \eqref{BRgenrz} we see that
this gives that $G_{12}=G_{22}=0$.
This means we have a rod $[\el^2,\infty]$ in the
direction $v=(0,1,0)$, i.e. in the $\partial / \partial x^2$
direction.
Using \eqref{deta} we see furthermore
that $x^2=\phi$ needs to have period
\begin{equation}
\label{phiper}
\Delta \phi = 2\pi \frac{\sqrt{1-b}}{1-c} \ ,
\end{equation}
to avoid a conical singularity for $r=0$ and $z\in [\el^2,\infty]$.
Since $v=-1$ is equivalent to $ R_1 - R_3 - (1+c)\el^2 = 0$
we see that this conical singularity corresponds to the one at
$v=-1$ mentioned above.
\end{itemize}
%

%%%%%%%%%%%%%%%%%%%%%%%%%%%%%%%%%%%%%%%%%%%%%%%%%%
\subsubsection*{Static black ring}

We consider here briefly the case of the static black ring,
obtained by setting $b=c$.
The static black ring was first discussed in \cite{Emparan:2001wk}.
The static black ring is in the class of generalized
Weyl solutions of \cite{Emparan:2001wk} since its metric is diagonal.

Putting $b=c$ in \eqref{genBRmet} one easily gets the
neutral black ring in C-metric coordinates. Note that $C=0$,
$G(\xi)= (1-\xi^2)F(\xi)$, $F(\xi) = 1 + c \xi$ and $0 < c < 1$.
Using \eqref{BRgenrz}-\eqref{e2nuBR}, we see that
the static black ring metric in
canonical coordinates
\eqref{themet}-\eqref{rdetG} takes the form \cite{Emparan:2001wk}
\begin{equation}
\label{BRstarz}
\begin{array}{c} \ds
G_{11} = - \frac{R_1+R_2-2c\el^2}{R_1+R_2+2c\el^2}
= - \frac{R_1 - z -c\el^2}{R_2 - z + c\el^2} \ ,
\\[4mm] \ds
G_{22} = R_3 - z+\el^2
\spa
G_{33} = \frac{(R_3 + z-\el^2)(R_2 - z+c\el^2)}{R_1 - z-c\el^2} \ ,
\\[4mm] \ds
e^{2\nu} = \frac{(R_1+R_2+2c\el^2)\left[ (1-c)R_1 + (1+c)R_2 + 2cR_3 \right]}{8(1-c^2) R_1 R_2 R_3} \ .
\end{array}
\end{equation}
The static black ring metric have previously been written
in canonical coordinates in \cite{Emparan:2001wk} since it falls
in the class of generalized Weyl solutions considered there.
The rod-structure of the static black ring is:
\begin{itemize}
\item The semi-infinite space-like rod $[-\infty,-c\el^2]$
in the $\partial / \partial x^3$ direction.
\item The finite time-like rod $[-c\el^2,c\el^2]$
in the $\partial / \partial x^1$ direction.
\item The finite space-like rod $[c\el^2,\el^2]$
in the $\partial / \partial x^3$ direction.
\item The semi-infinite space-like rod $[\el^2,\infty]$
in the $\partial / \partial x^2$ direction.
\end{itemize}
We see that all the rods are rectangular relative to each other.
The rod-structure of the static
black ring was previously described in \cite{Emparan:2001wk}.

%%%%%%%%%%%%%%%%%%%%%%%%%%
\subsubsection*{Getting the Myers-Perry black hole from the general
black ring solution}

We show here how one obtains the five-dimensional Myers-Perry rotating
black hole solution with one angular momentum, that we considered
in Section \ref{s:MPone}, from the general black ring solution.
This has previously been described in Ref.~\cite{Emparan:2004wy} in terms
of the C-metric coordinates used in the metric \eqref{genBRmet}.
Here we do it instead in terms of the canonical form of the metric
\eqref{BRgenrz}-\eqref{e2nuBR}.

We first note that we need to take the limit $c \rightarrow 1$,
since the $[c\el^2,\el^2]$ rod should be absent for the black
hole solution. By considering explicit expressions for the mass $M$ and
angular momentum $J_1$, plus the fact that $c \leq b < 1$,
one can see that $(1-b)/(1-c)$ and $\el^2/(1-c)$ should be fixed
as $c \rightarrow 1$. One can furthermore see that we can find
$\lambda$ and $\alpha$, so that
\begin{equation}
c = 1 - \epsilon \spa
b = 1 - \epsilon \cos^2 \lambda \spa
\el = \frac{\sqrt{\alpha}}{\cos \lambda} \sqrt{\epsilon} \ ,
\end{equation}
with the limit being defined as $\epsilon \rightarrow 0$.
Since $x^2$ and $x^3$ have periods \eqref{phiper} and \eqref{psiper}
we need to make the rescaling
\begin{equation}
x^2 = \frac{\cos \lambda}{\sqrt{\epsilon}} \tilde{x}^2 \spa
x^3 = \frac{\cos \lambda}{\sqrt{\epsilon}} \tilde{x}^3 \ ,
\end{equation}
so that now $\tilde{x}^2$ and $\tilde{x}^3$ have period $2\pi$
for $\epsilon \rightarrow 0$.
From the definition of the canonical $(r,z)$ coordinates, we see
that this means we should make the rescaling
\begin{equation}
r = \frac{\epsilon}{\cos^2 \lambda} \tilde{r}
\spa
z = \frac{\epsilon}{\cos^2 \lambda} \tilde{z} \ .
\end{equation}
This gives that
$\sqrt{r^2+(z \pm \el^2)^2} =\epsilon \sqrt{\tilde{r}^2+(\tilde{z} \pm \alpha)^2}
/\cos^2 \lambda$. Using this with the metric
\eqref{BRgenrz}-\eqref{e2nuBR} it is easy to see that
one gets the metric \eqref{MPcan} of
a Myers-Perry rotating black hole with
one angular momentum.

%%%%%%%%%%%%%%%%%%%%%%%%%%%%%%%%%%%%%%%%%%%%%%%%%%
\subsection{Regular rotating black ring}
\label{s:rotBR}

We now consider the regular black ring solution.
In Section \ref{s:genBR} we cured the conical singularities
at the $[-\infty,-c\el^2]$ and $[\el^2,\infty]$ rods
by imposing $x^2=\phi$ to have period \eqref{phiper} and $x^3=\psi$
to have period \eqref{psiper}.
However, we did not fix the potential conical singularity
at the $[c\el^2,\el^2]$ rod.
To ensure regularity at the $[c\el^2,\el^2]$ rod, $x^3=\psi$ should
have period \eqref{extrapsi}, which means
we need to impose
\begin{equation}
\label{bofc}
b = \frac{2c}{1+c^2} \ .
\end{equation}
Therefore, with \eqref{bofc} imposed, and with $x^2=\phi$ and $x^3=\psi$ having
their periods given by
\begin{equation}
\label{pp2}
\Delta \phi = \Delta \psi = \frac{2\pi}{\sqrt{1+c^2}} \ ,
\end{equation}
the rotating black ring solution \eqref{genBRmet} is regular
\cite{Emparan:2001wn,Emparan:2004wy}.
Note that the constant $C$ in \eqref{conC} now takes the form
\begin{equation}
C = \frac{2 c (1+c)}{1+c^2} \sqrt{\frac{1+c}{1-c}} \ .
\end{equation}
From \eqref{BRgenrz} we get that the
regular rotating black ring metric in
canonical coordinates
\eqref{themet}-\eqref{rdetG} is given by
\begin{equation}
\label{BRregrz}
\begin{array}{c} \ds
G_{11} = - \frac{(1+c) R_1 + (1-c)R_2 - 2c R_3 - 4c\el^2}{(1+c)
R_1 + (1-c)R_2 - 2c R_3 + 4c\el^2} \ ,
\\[4mm] \ds
G_{12} = - \frac{4c\kappa\sqrt{1+c}}{\sqrt{1-c}} \frac{R_3-R_1 +
(1+c)\el^2}{(1+c) R_1 + (1-c)R_2 - 2c R_3 + 4c\el^2} \ ,
\\[4mm] \ds
G_{33} = \frac{(R_3 + z-\el^2)(R_2 - z+c\el^2)}{R_1 - z-c\el^2} \
,
\\[4mm] \ds
e^{2\nu} = \left[ (1+c) R_1 + (1-c)R_2 - 2c R_3 + 4c \el^2 \right]
\frac{(1-c)R_1 + (1+c)R_2 + 2cR_3}{8(1-c^4) R_1 R_2 R_3} \ .
\end{array}
\end{equation}
One can furthermore find $G_{22}$ using \eqref{G22BR}.

%%%%%%%%%%%%%%%%%%%%%%%%%%%%%
\subsubsection*{Rod-structure}

The rod-structure of the regular rotating black ring solution
is easily obtained from the rod-structure of the general black ring
solution analyzed in Section \ref{s:genBR} by imposing \eqref{bofc}.
We list here therefore only a short summary of the rod-structure
of the regular rotating black ring:
\begin{itemize}
\item The semi-infinite space-like rod $[-\infty,-c\el^2]$.
This rod is in the direction $v=(0,0,1)$,
i.e. in the $\partial / \partial x^3$ direction.
\item The finite time-like rod $[-c\el^2,c\el^2]$.
This rod is in the direction
\begin{equation}
\label{vrotBR}
v = (1,\Omega,0)
\spa
\Omega = \frac{1}{2\kappa} \sqrt{\frac{1-c}{1+c}} \ .
\end{equation}
\item The finite space-like rod $[c\el^2,\el^2]$.
This rod is in the direction $v=(0,0,1)$,
i.e. in the $\partial / \partial x^3$ direction.
\item The semi-infinite space-like rod $[\el^2,\infty]$.
This rod is in the direction $v=(0,1,0)$,
i.e. in the $\partial / \partial x^2$ direction.
\end{itemize}

%%%%%%%%%%%%%%%%%%%%%%%%%%%%%
\subsubsection*{Asymptotic region}

For the regular
rotating black ring solution \eqref{BRregrz}
in the asymptotic region $\sqrt{r^2+z^2}\rightarrow \infty$
with $z/\sqrt{r^2+z^2}$ finite, we find
\begin{equation}
\begin{array}{c} \ds
G_{11} = - 1 + \frac{4c\el^2}{1-c} \frac{1}{\sqrt{r^2+z^2}}
+\CO((r^2+z^2)^{-1}) \ ,
\\[4mm] \ds
G_{12} = - 2 c \el^3 \left( \frac{1+c}{1-c} \right)^{\frac{3}{2}}
\frac{\sqrt{r^2+z^2}-z}{r^2+z^2} +\CO((r^2+z^2)^{-1}) \ ,
\\[4mm] \ds
G_{22} = \left( \sqrt{r^2+z^2} - z \right) \left[
1 + \frac{(1+c+2c^2)\el^2}{1-c} \frac{1}{\sqrt{r^2+z^2}}
+ \CO( (r^2+z^2)^{-1} ) \right] \ ,
\\[4mm] \ds
G_{33} = \left( \sqrt{r^2+z^2} + z \right) \left[
1 + \frac{(2c-1)\el^2}{\sqrt{r^2+z^2}}
+ \CO( (r^2+z^2)^{-1} ) \right] \ .
\end{array}
\end{equation}
Using \eqref{altasympt} we then get
\begin{equation}
M = \frac{3\pi c \el^2}{(1-c)(1+c^2)}
\spa
J_1 = 2\pi c \el^3 \left( \frac{1+c}{(1-c)(1+c^2)} \right)^{\frac{3}{2}}
\spa
\eta = \frac{3\pi \el^2(1-c+2c^2)}{2(1-c)(1+c^2)} \ ,
\end{equation}
along with $J_2 = 0$ and $\zeta = 0$,
where we used that $\varepsilon = 1/\sqrt{1+c^2}$ from \eqref{pp2}.
Note that
\begin{equation}
\frac{J_1^2}{M^3} = \frac{4(1+c)^3}{27\pi c} \ .
\end{equation}
We see that this has the minimum at $c=1/2$ with value $1/\pi$.
For $c\rightarrow 0$ it goes to infinity, while for
$c = 1$ it has the value $32/(27\pi)$. This is in accordance with
\cite{Emparan:2001wn,Elvang:2003yy,Emparan:2004wy}.

%%%%%%%%%%%%%%%%%%%%%%%%%%%%%%%%%%%%%%%%%%%%%%%%%%%%%%%%%%%%%%
\section{Discussion and conclusions}
\label{s:concl}

The main results of this paper are as follows.
We found in Section \ref{secstataxi}
that the metric of stationary and axisymmetric
pure gravity solutions in $D$ dimensions can be written in
the form (see Eqs.~\eqref{themet}-\eqref{rdetG})
\[
ds^2 = \sum_{i,j=1}^{D-2} G_{ij} dx^i dx^j
+ e^{2\nu} (dr^2+dz^2) \spa
r^2 = | \det G | \ ,
\]
apart from a subclass of solutions with constant $\det G$ that
is considered in Appendix \ref{appdetconst}.
The equation on the $D-2$ by $D-2$ dimensional symmetric matrix $G$
was found to take the simple form (see Eqs.~\eqref{formalG})
\[
G^{-1} \grad^2 G = ( G^{-1} \grad G )^2 \ ,
\]
where $\grad$ is the gradient on a three-dimensional
flat Euclidean space, with metric \eqref{unmet}.
The function $\nu$ can then be found from $G$ using the
integrable equations \eqref{nueqs}.

In Section \ref{s:nearrod} we considered then the behavior of
$G$ for $r \rightarrow 0$. We generalized the concept of
rods of \cite{Emparan:2001wk} so that it can be used for
stationary and axisymmetric solutions. One of the key
points is that for each rod $[z_1,z_2]$ one has a direction
in the $(D-2)$-dimensional vector space spanned
by the Killing vector fields.

In Section \ref{s:asymp} we analyzed the asymptotic region
of four- and five-dimensional asymptotically flat solutions.
In particular we identified how to
read off the asymptotic quantities.

Finally, in Sections \ref{s:rotBH} and \ref{s:BRsol} we wrote down the metrics
of the five-dimensional rotating black hole of Myers and Perry
and the rotating black ring of Emparan and Reall in the
canonical form \eqref{themet}-\eqref{rdetG}. Furthermore,
we analyzed the structure of the rods according to Section \ref{s:nearrod}
and moreover the asymptotic region according to Section \ref{s:asymp}.

The results of this paper have at least three
interesting applications:
\begin{itemize}
\item
Finding new stationary and axisymmetric solutions
using the canonical form of the metric and the Einstein equations.
For example one can look for new five-dimensional black ring solutions
with two angular momenta, or for
new solutions with a rotating black hole attached to
a Kaluza-Klein bubble, as advocated in \cite{Elvang:2004iz}.
\item
Understanding the rod-structure of known stationary
and axisymmetric solutions.
\item
Understanding better the uniqueness properties for
higher-dimensional black holes.
In four dimensions, the Carter-Robinson uniqueness theorem
\cite{Carter:1971,Robinson:1975}
on the Kerr rotating black hole rests on
using the Papapetrou form \eqref{papmet} of the metric.
We expect therefore similar arguments to be applicable
in higher dimensions, although they of course cannot prove
any kind of strict uniqueness for five-dimensional rotating
black holes due to the existence of rotating black rings.
\end{itemize}

%%%%%%%%%%%%%%%%%%%%%%%%%%%%%%%%%%%%%%%%%%%%%%%%%%%%%%%%%%%%%%
\section*{Acknowledgments}

We thank H. Elvang, N. Obers and P. Olesen for discussions
and comments.

\begin{appendix}
%%%%%%%%%%%%%%%%%%%%%%%%%%%%%%%%%%%%%%%%%%%%%%%%%%%%%%%%%%%%%%
\section{Special class of solutions}
\label{appdetconst}

In this appendix we consider stationary and axisymmetric
solutions to the vacuum Einstein equations for which
$\det (G_{ij})$ is constant, with $G_{ij}$ defined by Eq.~\eqref{met1}.

We can always find coordinates $(r,z)$ so that the metric
\eqref{met1} can be written
\begin{equation}
\label{mett}
ds^2 = \sum_{i,j=1}^{D-2} G_{ij} dx^i dx^j +
e^{2\nu} ( dr^2 + dz^2 ) \ .
\end{equation}
This is possible since any two-dimensional manifold is conformally flat.
The metric \eqref{mett} is obviously the same as \eqref{themet}.
However, the important difference is that the constraint \eqref{rdetG}
is replaced by restricting $\det( G_{ij} )$ to be constant.

Note first that demanding $\det( G_{ij} )$ to be constant
leads to the identities
\begin{equation}
\sum_{i,j=1}^{D-2} G^{ij} \partial_a G_{ij} = 0 \spa
\sum_{i,j=1}^{D-2} G^{ij} \partial_a \partial_b G_{ij}
= \sum_{i,j,k,l=1}^{D-2} G^{ij} \partial_a G_{jk} G^{kl} \partial_b G_{li}
\ ,
\end{equation}
with $a,b=r,z$.
Computing the Ricci tensor for the metric \eqref{mett} and using
the constraint that $\det (G_{ij})$ is constant, we get that
the vacuum Einstein equations can be written
\begin{equation}
\label{Geqs}
\begin{array}{c} \ds
(\partial_r^2 + \partial_z^2 ) G_{ij} =
\sum_{k,l=1}^{D-2} \partial_r G_{ik} G^{kl} \partial_r G_{lj}
+ \sum_{k,l=1}^{D-2} \partial_z G_{ik} G^{kl} \partial_z G_{lj} \ ,
\\[4mm] \ds
\sum_{i,j,k,l=1}^{D-2} G^{ij} \partial_r G_{jk} G^{kl} \partial_r G_{li}
= \sum_{i,j,k,l=1}^{D-2} G^{ij} \partial_z G_{jk} G^{kl} \partial_z G_{li}
\\[4mm] \ds
\sum_{i,j,k,l=1}^{D-2}
G^{ij} \partial_r G_{jk} G^{kl} \partial_z G_{li} = 0
\end{array}
\end{equation}
\begin{equation}
\label{nueqq}
(\partial_r^2 + \partial_z^2 ) \nu =
- \frac{1}{8} \sum_{i,j,k,l=1}^{D-2} \left( G^{ij} \partial_r G_{jk} G^{kl} \partial_r G_{li} + G^{ij} \partial_z G_{jk} G^{kl} \partial_z G_{li} \right) \ .
\end{equation}
In conclusion, we can find solutions in the form of \eqref{mett}
with $\det (G_{ij})$ being constant by first finding
a $G_{ij} (r,z)$ solving \eqref{Geqs} and
then finding a solution for $\nu(r,z)$ of \eqref{nueqq}.

%%%%%%%%%%%%%%%%%%%%%%%%%%%%
\subsubsection*{Four-dimensional examples}

In four dimensions, we have a well-known class of solutions to
Eqs.~\eqref{Geqs}-\eqref{nueqq} in the form of a particular kind
of pp-wave solutions.
These pp-wave solutions have
\begin{equation}
G_{11} = - 1 - H(r,z)
\spa
G_{22} = 1 - H(r,z)
\spa
G_{12} = - H(r,z) \ .
\end{equation}
We see immediately that $\det ( G_{ij} ) = -1 $.
Furthermore, one can check that the Eqs.~\eqref{Geqs} with $D=4$
are solved, provided $H(r,z)$ obeys
\begin{equation}
\label{HHeq}
\left( \partial_r^2 + \partial_z^2 \right) H(r,z) = 0 \ .
\end{equation}
Finally, $\nu(r,z) = 0$ solves Eq.~\eqref{nueqq}. Therefore,
the pp-wave metrics
\begin{equation}
ds^2 = - dt^2 + dx^2 - H (dt+dx)^2  + dr^2 + dz^2 \ ,
\end{equation}
with $H(r,z)$ obeying Eq.~\eqref{HHeq},
are in the class of solutions described by the metric
\eqref{mett} with $\det (G_{ij})$ being constant~\cite{Stephani:2003}.
Note moreover that any $\nu(r,z)$ solving
$(\partial_r^2 + \partial_z^2) \nu = 0$ also gives a solution.

%%%%%%%%%%%%%%%%%%%%%%%%%%%%%%%%%%%%%%%%%%%%%%%%%%%%%%%%%%%%%%
\section{Analysis of $\det ( G_{ij} )$}
\label{appdetG}

In this appendix we study the behavior of $\det ( G_{ij} )$ as a function.
If we start with the metric \eqref{met1} we can always put it in the
form
\begin{equation}
\label{newmet1}
ds^2 = \sum_{i,j=1}^{D-2} G_{ij} dx^i dx^j
+ C ( du^2 + dv^2 ) \ ,
\end{equation}
where $C(u,v)$ and $G_{ij} (u,v)$ are functions only
of $u$ and $v$.
That we can bring the metric
\eqref{met1} to this form is easily seen from the fact that
any two-dimensional manifold is conformally flat.
Define now
\begin{equation}
f = \sqrt{ | \det ( G_{ij} ) |}  \ .
\end{equation}
In the following we study the function $f(u,v)$.
By computing the Ricci tensor for the metric \eqref{newmet1}
we get
\begin{equation}
\sum_{i,j=1}^{D-2} G^{ij} R_{ij} = - \frac{1}{C f}
\left( \frac{\partial^2}{\partial u^2} + \frac{\partial^2}{\partial v^2}
\right) f  \ .
\end{equation}
Now, since we consider solutions that are Ricci flat, we get
that
\begin{equation}
\left( \frac{\partial^2}{\partial u^2} + \frac{\partial^2}{\partial v^2}
\right) f = 0  \ .
\end{equation}
If we define the complex variable $\omega = u + i v$, along
with the derivatives
$\partial = \frac{\partial}{\partial u} + i \frac{\partial}{\partial v}$ and
$\bar{\partial} = \frac{\partial}{\partial u} - i \frac{\partial}{\partial v}$,
we see that $\bar{\partial} \partial f = 0$.
Therefore, $\partial f$ is a holomorphic function.
We know from elementary complex analysis (see for
example \cite{Flanigan:1972}) that
either the zeroes of a holomorphic function are isolated
or the function is identically zero
(assuming the set that the function is defined on
is simply connected).
Since $\partial f = \frac{\partial f}{\partial u}
+ i \frac{\partial f}{\partial v}$
we can draw the conclusion:
\begin{itemize}
\item Either $f (u,v) $ is a constant function or
$(\frac{\partial f}{\partial u} , \frac{\partial f}{\partial v}) \neq (0,0)$ except
in isolated points.
\end{itemize}

%%%%%%%%%%%%%%%%%%%%%%%%%%%%%%%%%%%%%%%%%%%%%%%%%%%%%%%%%%%%%%
\section{Diagonalizing a two-dimensional metric}
\label{2Dmet}

In this appendix we prove, for the sake of clarity
and completeness, the rather basic result that given a
well-behaved function on a two-dimensional Riemannian manifold
one can diagonalize the metric with the given function being
one of the coordinates.

Consider a two-dimensional Riemannian manifold $M$ with a coordinate system
$(y^1,y^2)$.
Write the metric as
\begin{equation}
ds^2 = \hat{g}_{ab} dy^a dy^b \ .
\end{equation}
Let $z^1(y^1,y^2)$ be a given function
with
$( \frac{\partial z^1}{\partial y^1},\frac{\partial z^1}{\partial y^2})
\neq (0,0)$.
We want to show that we can find a function $z^2(y^1,y^2)$ so
that $(z^1,z^2)$ is a new coordinate system and so
that the metric in $z^a$ coordinates is diagonal, i.e. so that
$g_{12} = 0$, where we write the metric
as $ds^2 = g_{ab} dz^a dz^b$.
Equivalently, we can demand that $g^{12}=0$. This is the same as
\begin{equation}
\label{gabup}
\hat{g}^{ab} \frac{\partial z^1 }{\partial y^a}
\frac{\partial z^2}{\partial y^b} = 0 \ .
\end{equation}
Now, define the vector field $V= V^1 \frac{\partial}{\partial y^1}
+ V^2 \frac{\partial}{\partial y^2}$ by
\begin{equation}
V^a = \hat{g}^{ab} \frac{\partial z^1 }{\partial y^b} \ .
\end{equation}
Consider now the integral curves of $V$.
Define an equivalence relation $\sim$ on $M$ where
two points $p,q \in M$ are equivalent, i.e.
$p \sim q$, if they are connected by an integral curve.
Then we can define the quotient space $M/\sim$.
Clearly, $M/\sim$ is a one-dimensional space.
Let now $z^2$ be a coordinate on $M/\sim$.
We then extend the scalar field $z^2$ on
$M/\sim$ to a scalar field $z^2$ on $M$.
Clearly, this scalar field $z^2$ on $M$ has the property
that $z^2$ is constant on the integral curves of $V$.
Since $z^2$ is constant on the integral curves of $V$ we get that
\begin{equation}
V^a \frac{\partial z^2}{\partial y^a} = 0 \ ,
\end{equation}
which is the same as \eqref{gabup}.
We have therefore proven that for any given function $z^1(y^1,y^2)$ with
$(\frac{\partial z^1}{\partial y^1},\frac{\partial z^1}{\partial y^2})
\neq (0,0)$
we can find a function $z^2(y^1,y^2)$ such that
\begin{equation}
ds^2 = \hat{g}_{ab} dy^a dy^b
= A (dz^1)^2 + B (dz^2)^2 \ ,
\end{equation}
and so that $(z^1,z^2)$ is a coordinate system on the two-dimensional
manifold.

%%%%%%%%%%%%%%%%%%%%%%%%%%%%%%%%%%%%%%%%%%%%%%%%%%%%%%%%%%%%%%
\section{Computation of Ricci tensor}
\label{appricci}

\subsubsection*{Computation of the Ricci tensor with general $\Lambda$}

We consider first the $D$-dimensional metric
\begin{equation}
\label{ddimmet}
ds^2 = \sum_{i,j=1}^{D-2} G_{ij} dx^i dx^j +
e^{2\nu} ( dr^2 + \Lambda \, dz^2 ) \ ,
\end{equation}
with
\begin{equation}
r = \sqrt{ | \det ( G_{ij} ) | } \ ,
\end{equation}
where $G_{ij}$, $\nu$ and $\Lambda$ are functions of $r$ and $z$
only.
The non-zero components
of the Christoffel symbols for the metric \eqref{ddimmet} are
\begin{equation}
\begin{array}{c}  \ds
\Gamma^r_{ij} = - \frac{1}{2} e^{-2\nu} \partial_r G_{ij}
\spa
\Gamma^z_{ij} = - \frac{1}{2} e^{-2\nu} \Lambda^{-1} \partial_z G_{ij} \ ,
\\[4mm] \ds
\Gamma^i_{rj}
= \frac{1}{2} \sum_{k=1}^{D-2} G^{ik} \partial_r G_{jk}
\spa
\Gamma^i_{zj}
= \frac{1}{2} \sum_{k=1}^{D-2} G^{ik} \partial_z G_{jk} \ ,
\\[4mm] \ds
\Gamma^r_{rr} = \partial_r \nu \spa
\Gamma^z_{zz} = \partial_z \nu + \frac{1}{2\Lambda} \partial_z \Lambda \spa
\Gamma^r_{rz} = \partial_z \nu \spa
\Gamma^z_{zr} = \partial_r \nu + \frac{1}{2\Lambda} \partial_r \Lambda  \ ,
\\[4mm] \ds
\Gamma^r_{zz} = - \Lambda \partial_r \nu -  \frac{1}{2C} \partial_r \Lambda \spa
\Gamma^z_{rr} = - \frac{1}{\Lambda} \partial_z \nu \ .
\end{array}
\end{equation}
Note now that since $r = \sqrt{|\det G_{ij} |}$
we have
\begin{equation}
\sum_{i,j=1}^{D-2} G^{ij} \partial_r G_{ij} = \frac{2}{r}
\spa
\sum_{i,j=1}^{D-2} G^{ij} \partial_z G_{ij} = 0 \ .
\end{equation}
Using this we get
\begin{equation}
\sum_{i,j=1}^{D-2} G^{ij} \Gamma^r_{ij} = - \frac{1}{r} e^{-2\nu}
\spa
\sum_{i,j=1}^{D-2} G^{ij} \Gamma^z_{ij} = 0
\spa
\sum_{i=1}^{D-2} \Gamma^i_{ri} = \frac{1}{r}
\spa
\sum_{i=1}^{D-2} \Gamma^i_{zi} = 0  \ .
\end{equation}
We compute then
\begin{equation}
\label{Rijlamb}
\begin{array}{rcl}
2 e^{2\nu} R_{ij} &=& \ds
 - \partial_r^2 G_{ij}
- \frac{1}{r} \partial_r G_{ij}
- \frac{\partial_r \Lambda}{2\Lambda} \partial_r G_{ij}
- \frac{1}{\Lambda} \partial_z^2 G_{ij}
+ \frac{\partial_z \Lambda}{2\Lambda^2} \partial_z G_{ij}
\\[4mm] && \ds
+ \sum_{k,l=1}^{D-2} G^{kl} \partial_r G_{ki} \partial_r G_{lj}
+ \frac{1}{\Lambda} \sum_{k,l=1}^{D-2}
G^{kl} \partial_z G_{ki} \partial_z G_{lj}   \ .
\end{array}
\end{equation}
Notice now that from the fact that $r = \sqrt{|\det G_{ij} |}$ we have
\begin{equation}
\label{ff}
\begin{array}{c} \ds
- \sum_{i,j=1}^{D-2} G^{ij} \partial_r^2 G_{ij}
+ \sum_{i,j,k,l=1}^{D-2}
G^{ij} G^{kl}
\partial_r G_{ki} \partial_r G_{lj} = \frac{2}{r^2}   \ ,
\\[5mm] \ds
- \sum_{i,j=1}^{D-2} G^{ij} \partial_z^2 G_{ij}
+  \sum_{i,j,k,l=1}^{D-2}  G^{ij}
G^{kl} \partial_z G_{ki} \partial_z G_{lj}
= 0  \ .
\end{array}
\end{equation}
Using \eqref{ff} together with \eqref{Rijlamb}, we get
\begin{equation}
\label{trGR}
\sum_{i,j=1}^{D-2} G^{ij} R_{ij}
= - \frac{\partial_r \Lambda}{2 e^{2\nu} \Lambda r}   \ .
\end{equation}
%

%%%%%%%%%%%%%%%%%%%%%%%%%%
\subsubsection*{Computation of the Ricci tensor with $\Lambda=1$}

We now set $\Lambda = 1$ in the metric \eqref{ddimmet}.
The non-zero components of the Ricci tensor can then
be computed to be
\begin{equation}
\label{RicciComp}
\begin{array}{c} \ds
2 e^{2\nu} R_{ij} = -
\left( \partial_r^2 + \frac{1}{r} \partial_r + \partial_z^2 \right) G_{ij}
+ \sum_{k,l=1}^{D-2}
 G^{kl} \partial_r G_{ki} \partial_r G_{lj}
+ \sum_{k,l=1}^{D-2}
G^{kl} \partial_z G_{ki} \partial_z G_{lj}   \ ,
\\[4mm]  \ds
R_{rr} = - \partial_r^2 \nu - \partial_z^2 \nu
+ \frac{1}{r^2} + \frac{1}{r} \partial_r \nu
- \frac{1}{4} \sum_{i,j,k,l=1}^{D-2}  G^{ij}
G^{kl} \partial_r G_{ik} \partial_r G_{jl}   \ ,
\\[4mm]  \ds
R_{zz} = - \partial_r^2 \nu - \partial_z^2 \nu
- \frac{1}{r} \partial_r \nu
- \frac{1}{4} \sum_{i,j,k,l=1}^{D-2}  G^{ij}
G^{kl} \partial_z G_{ik} \partial_z G_{jl}   \ ,
\\[4mm]  \ds
R_{rz} = \frac{1}{r} \partial_z \nu
- \frac{1}{4} \sum_{i,j,k,l=1}^{D-2}  G^{ij}
G^{kl} \partial_r G_{ik} \partial_z G_{jl}   \ .
\end{array}
\end{equation}
%

%%%%%%%%%%%%%%%%%%%%%%%%%%%%%%%%%%%%%%%%%%%%%%%%%%%%%%%%%%%%%
\section{Properties of the equations for $G_{ij} (r,z)$}
\label{appprop}

In this appendix we derive several useful properties of the equations
\eqref{Gijeqs} for $G_{ij} (r,z)$. We use in the following the formal
rewriting of these equations in the form of Eq.~\eqref{formalG}.

Let $A(r,z)$ and $B(r,z)$ be $D-2$ times $D-2$ matrices obeying
\begin{equation}
\label{comAB}
[ A(r,z) , B(r',z') ] = 0 \ ,
\end{equation}
for any $(r,z)$ and $(r',z')$.
Note that this means that $A(r,z)$ or any derivative of $A(r,z)$
commutes with $B(r,z)$ or any derivative of $B(r,z)$.
Write now $G = AB$. Then we have
\begin{equation}
G^{-1} \grad^2 G - (G^{-1} \grad G)^2
= A^{-1} \grad^2 A - (A^{-1} \grad A)^2 +
B^{-1} \grad^2 B - (B^{-1} \grad B)^2  \ .
\end{equation}
From this equation we get the following lemma:
\begin{lemma}
\label{lemAB}
Let $A(r,z)$ and $B(r,z)$ be $D-2$ times $D-2$ matrices that
commutes as in \eqref{comAB}.
If $A$ and $B$ obey the differential equations
\begin{equation}
A^{-1} \grad^2 A = ( A^{-1} \grad A)^2 \spa
B^{-1} \grad^2 B = ( B^{-1} \grad B)^2  \ ,
\end{equation}
then the matrix $G = AB$ obeys
$G^{-1} \grad^2 G = ( G^{-1} \grad G)^2$.
\squ
\end{lemma}
The consequence of this lemma is that we can combine
solutions into new solutions, as long as \eqref{comAB} is obeyed.
An important use of lemma \ref{lemAB} is the following corollary:
\begin{corollary}
\label{corfA}
Let $A(r,z)$ be a $D-2$ times $D-2$ matrix and
let $f(r,z)$ be a function.
If $G$ and $f$ obey the differential equations
\begin{equation}
A^{-1} \grad^2 A = ( A^{-1} \grad A)^2 \spa
\grad^2 f = 0  \ ,
\end{equation}
then the matrix $G = e^f A$ obeys
$G^{-1} \grad^2 G = ( G^{-1} \grad G)^2$.
\squ
\end{corollary}
Another important situation where the lemma \ref{lemAB} can be applied and
where the implication of lemma \ref{lemAB} in fact can be reversed
is expressed in the following lemma:
\begin{lemma}
\label{lemdirect}
Let the $D-2$ times $D-2$ matrix $G(r,z)$ be such that
$G = A \oplus B $, where $A(r,z)$ is a $k$ times $k$ matrix
and $B(r,z)$ is a $D-2-k$ times $D-2-k$ matrix.
I.e. $G$ is the geometric direct sum of $A$ and $B$.
In this case, it is clear that
$A$ and $B$ obey the differential equations
\begin{equation}
A^{-1} \grad^2 A = ( A^{-1} \grad A)^2 \spa
B^{-1} \grad^2 B = ( B^{-1} \grad B)^2  \ ,
\end{equation}
if and only if $G$ obeys
$G^{-1} \grad^2 G = ( G^{-1} \grad G)^2$. \squ
\end{lemma}
This lemma can of course be used successively for the cases
where $G(r,z)$ can be split up to the direct sum of several
matrices acting on linearly independent subspaces,
i.e. $G = A_1 \oplus A_2 \oplus \cdots \oplus A_n$.
An important special case of this is when $G$ is diagonal.
We have the following corollary of lemma \ref{lemdirect}:
\begin{corollary}
\label{lemdiag}
Let the $D-2$ times $D-2$ matrix $G(r,z)$ be a diagonal
matrix
\begin{equation}
G = \diag (\pm \exp(2 U_1),\exp(2 U_2),...,\exp(2 U_{D-2}))  \ ,
\end{equation}
where $U_i(r,z)$, $i=1,...,D-2$, are functions.
Then
\begin{equation}
\grad^2 U_i = 0 \spa i =1,...,D-2  \ ,
\end{equation}
if and only if $G^{-1} \grad^2 G = ( G^{-1} \grad G)^2$.
\squ
\end{corollary}
When $G(r,z)$ is diagonal it corresponds to a generalized
Weyl solution \eqref{GWmet} (see \cite{Emparan:2001wk}),
since the $D-2$ Killing vector fields are orthogonal.
We see that \eqref{Gijeqs} correctly reduce to
\eqref{Ulap}. Moreover, it is clear that
$\det (G_{ij}) = \pm r^2$ is equivalent
to $\sum_{i=1}^{D-2} U_i = \log r$.

We have also a general result for the inverse of a matrix:
\begin{lemma}
\label{leminv}
An $n$ by $n$ invertible matrix $G(r,z)$ obeys the equation
$G^{-1} \grad^2 G = ( G^{-1} \grad G)^2$
if and only if the inverse matrix $G^{-1}$ obeys
the corresponding equation
\begin{equation}
G \grad^2 G^{-1} = ( G \grad G^{-1})^2  \ .
\end{equation}
\squ
\end{lemma}
This lemma can be used to find new solutions from already known solutions.
Of course, one has to remember that the complete $G$ matrix
moreover should have $|\det G | = r^2$.
The following corollary is one way to take this into account:
\begin{corollary}
Let $G(r,z)$ be a $D-2$ by $D-2$ matrix with $|\det G| = r^2$
and $G^{-1} \grad^2 G = ( G^{-1} \grad G)^2$.
Then the matrix $M = r^{\frac{4}{D-2}} G^{-1}$ obeys
$\det M = \det G$ and $M^{-1} \grad^2 M = ( M^{-1} \grad M)^2$.
\squ
\end{corollary}

We note that we can multiply constant matrices on
solutions:
\begin{lemma}
\label{lemmult}
Let the $D-2$ by $D-2$ matrix $G(r,z)$ solve the equation
$G^{-1} \grad^2 G = ( G^{-1} \grad G)^2$
and let $A$ and $B$ be constant invertible matrices.
Then the matrix $M = A G B$ obeys
$M^{-1} \grad^2 M = ( M^{-1} \grad M)^2$.
\squ
\end{lemma}

Finally, an important and useful theorem that concerns systems with
an orthogonal Killing vector is the following:
\begin{theorem}
\label{theored}
Consider the class of metrics with $G_{1i} = 0$, $i=2,...,D-2$,
i.e. with the Killing vector $V_{(1)}= \frac{\partial}{\partial x^1}$
being orthogonal to the $D-3$ other Killing vector fields.
Then we can always write $G$ as
\begin{equation}
\label{GUM}
G = s \, e^{2U} \oplus e^{-\frac{2}{D-3}U} M
= \left( \begin{array}{cccc} s \, e^{2U} & 0 & \cdots & 0 \\
0 & \ & \ & \ \\
\vdots & \ & e^{-\frac{2}{D-3}U} M & \ \\
0 & \ & \ & \
\end{array} \right)   \ ,
\end{equation}
where $s= \pm 1$, $U(r,z)$ is a function and $M(r,z)$ is
a $D-3$ by $D-3$ symmetric real matrix
with $\det M = s \det G$ so that $| \det M | = r^2$.
Moreover, $G$ obeys $G^{-1} \grad^2 G = (G^{-1} \grad G)^2$
if and only if
\begin{equation}
\label{UMeq}
\grad^2 U = 0 \spa
M^{-1} \grad^2 M = (M^{-1} \grad M)^2  \ .
\end{equation}
\proof
It is trivial to see that we can always write $G$ on the form \eqref{GUM}.
That $G^{-1} \grad^2 G = (G^{-1} \grad G)^2$ if and only \eqref{UMeq}
is true follows from using lemma \ref{lemdirect} together
with lemma \ref{corfA}.
\squ
\end{theorem}
Theorem \ref{theored} is useful
since it allows one to take a $D=n$ dimensional solution and creating
new non-trivial $D=n+1$ dimensional solutions. Moreover,
for a $D=n$ dimensional solution with a Killing vector orthogonal
to all other $D-3$ Killing vector fields we can reduce the system
of equations to be a 3-dimensional Laplace equation
together with, but decoupled from,
the equations for a $D=n-1$ dimensional solution.

%%%%%%%%%%%%%%%%%%%%%%%%%%%%%%%%%%%%%%%%%%%%%%%%%%%%%%%%%%%%%%
\section{Singularities at $r=0$}
\label{appsing}

We consider in this appendix what happens
for solutions that have more than one eigenvalue of
$G(r,z)$ going to zero for $r\rightarrow 0$.
We restrict for simplicity here to the case with two eigenvalues
going to zero for $r\rightarrow 0$. One can easily extend the
argument to consider more than two eigenvalues.

We begin by considering the solution
\begin{equation}
\label{D4sys}
G_{11} = r^{2a} \spa
G_{22} = r^{2-2a} \spa
e^{2\nu} = r^{-2a(1-a)} \ ,
\end{equation}
where $0 \leq a \leq 1$. This solves the equations
\eqref{Gijeqs}-\eqref{nueqs}. We compute the curvature invariant
\begin{equation}
R_{\mu\nu\rho\sigma} R^{\mu\nu\rho\sigma}
= 16 a^2 (1-a)^2 (1-a+a^2) r^{-4(1-a+a^2)} \ .
\end{equation}
Since $1-a+a^2$ is strictly positive we see that there is a
curvature singularity at $r=0$, unless $a=0$ or $a=1$ which in both cases
corresponds to only having one eigenvalue going to zero.
Therefore, for this solution we see that having two eigenvalues
going to zero invariably leads to a curvature singularity.

Consider now a general solution for which we have an interval
$[z_1,z_2]$
so that for any $z \in [z_1,z_2]$ we have that
two eigenvalues of $G(r,z)$ going to zero for $r \rightarrow 0$.
Then using the same type of arguments as in Section \ref{s:rods}
we can make a constant orthogonal transformation of $G(r,z)$
so that $G_{1i} (0,z) = G_{2i}(0,z) = 0$
for $i=1,2,...,D-2$,
for a given $z_1 < z < z_2$.
Therefore, given the structure of the equations
\eqref{Gijeqs}-\eqref{nueqs}, we see that we effectively
can reduce this system to the example given by Eq.~\eqref{D4sys}.
In conclusion, we have shown that for any solution having two eigenvalues
of $G(r,z)$ that go to zero as $r\rightarrow 0$ for a given $z$,
we get
curvature singularities, except perhaps in isolated points
on the $z$-axis
corresponding to the endpoints of the interval given above.
As mentioned above, one can easily extend these arguments to consider
more than two eigenvalues going to zero.

%%%%%%%%%%%%%%%%%%%%%%%%%%%%%%%%%%%%%%%%%%%%%%%%%%%%%%%%%%%%%%
\section{Prolate spherical coordinates}
\label{a:prolate}

We define here the {\sl prolate spherical coordinates} $(x,y)$
which for certain stationary and axisymmetric solutions
are convinient to use.
The prolate spherical coordinates were introduced for four-dimensional
stationary and axisymmetric solutions in \cite{Zipoy:1966}
(see also \cite{Stephani:2003,Chandrasekhar:1992,Heusler:1996}).
The prolate spherical coordinates are used
to describe rotating black hole solutions in Section \ref{s:rotBH}.

The prolate spherical coordinates $(x,y)$ are defined in
terms of the canonical $(r,z)$ coordinates by
\begin{equation}
\label{defprol}
r = \alpha \sqrt{(x^2 - 1) (1 - y^2)}
\spa
z = \alpha \, x \, y \ ,
\end{equation}
where $\alpha > 0$ is a constant.
We take $x$ and $y$ to have the ranges
\begin{equation}
x \geq 1 \spa
-1 \leq y \leq 1 \ .
\end{equation}
We have
\begin{equation}
dr^2 + dz^2 = \alpha^2 (x^2 - y^2)
\left[ \frac{dx^2}{x^2-1} + \frac{dy^2}{1-y^2} \right] \ .
\end{equation}
Note that Eqs.~\eqref{Gijeqs} can be written
in prolate spherical coordinates as
\begin{equation}
\partial_x \left[ (x^2-1) \partial_x G \right]
+ \partial_y \left[ (1-y^2) \partial_y G \right]
= (x^2-1) (\partial_x G) G^{-1} \partial_x G
+ (1-y^2) (\partial_y G) G^{-1} \partial_y G \ .
\end{equation}
We now give the transformation from $(x,y)$ coordinates to $(r,z)$
coordinates.
Defining
\begin{equation}
R_+ = \sqrt{r^2 + (z + \alpha)^2} \spa R_- = \sqrt{r^2 + (z - \alpha)^2} \ ,
\end{equation}
one can easily check using \eqref{defprol} that
\begin{equation}
R_+ = \alpha ( x + y ) \spa
R_- = \alpha ( x - y ) \ .
\end{equation}
Therefore, we see that
\begin{equation}
\label{xyfrz}
x = \frac{R_+ + R_-}{2\alpha } \spa
y = \frac{R_+ - R_-}{2\alpha } \ .
\end{equation}
Furthermore, we note that
\begin{equation}
R_\pm + z \pm \alpha
= \alpha (x \pm 1 )( 1+ y)
\spa
R_\pm - (z \pm \alpha)
= \alpha (x \mp 1 )( 1- y) \ .
\end{equation}

If we consider the asymptotic region $\sqrt{r^2+z^2} \rightarrow \infty$
with $z/\sqrt{r^2+z^2}$ finite,
we see that
\begin{equation}
\label{xyrzasymp}
x \simeq \frac{1}{\alpha} \sqrt{r^2+z^2}
\spa
y \simeq \frac{z}{ \sqrt{r^2+z^2}} \ .
\end{equation}
Thus, the asymptotic region in terms of
the prolate spherical coordinates is $x \rightarrow \infty$
and $y$ being finite.

%%%%%%%%%%%%%%%%%%%%%%%%%%%%%%%%%%%%%%%%%%%%%%%%%%
\section{C-metric coordinates}
\label{s:cmet}

We consider in this section the coordinate transformation
from general C-metric coordinates to the canonical $(r,z)$
coordinates.%
\footnote{See \cite{Pravda:2000vh} for a review of the C-metric and
the transformation from C-metric coordinates to
canonical $(r,z)$ coordinates in the context of four-dimensional
Weyl solutions.}
The general C-metric coordinates $(u,v)$ are here defined
in relation to $(r,z)$ by
\begin{equation}
\label{genrzuv}
\begin{array}{c} \ds
r = \frac{2\el^2 \sqrt{-G(u)G(v)}}{(u-v)^2}  \ ,
\\[4mm] \ds
e^{2\nu} ( dr^2 + dz^2 )
= \zeta(u,v) \left( \frac{du^2}{G(u)} - \frac{dv^2}{G(v)} \right) \ ,
\end{array}
\end{equation}
where $\el$ is a constant, and
$\zeta(u,v)$ and $G(\xi)$ are functions that depend
on the particular solution that we consider.
The goal is now to find the $z$ coordinate.
From $g^{rz} = 0$ we find
\begin{equation}
\label{eqq1}
\frac{\partial z}{\partial v} = - \frac{g_{vv}}{g_{uu}}
\frac{\partial r}{\partial u}
\left( \frac{\partial r}{\partial v} \right)^{-1}
\frac{\partial z}{\partial u} \ .
\end{equation}
Using this together with $g_{uv} = 0$ one can easily derive that
\begin{equation}
\frac{\partial z}{\partial u} = s \sqrt{ \frac{g_{uu}}{g_{vv}} }
\frac{\partial r}{\partial v}
\spa
\frac{\partial z}{\partial v} = - s \sqrt{ \frac{g_{vv}}{g_{uu}} }
\frac{\partial r}{\partial u}
\spa
s = \pm 1 \ .
\end{equation}
This gives
\begin{equation}
\label{partz}
s \frac{\partial z}{\partial u} = - \frac{\el^2 G'(v)}{(u-v)^2}
- \frac{4 \el^2 G(v)}{(u-v)^3}
\spa
s \frac{\partial z}{\partial v} = - \frac{\el^2 G'(u)}{(u-v)^2}
+ \frac{4 \el^2 G(u)}{(u-v)^3} \ .
\end{equation}
The integrability condition
\begin{equation}
\frac{\partial}{\partial v} \frac{\partial z}{\partial u}
= \frac{\partial}{\partial u} \frac{\partial z}{\partial v} \ ,
\end{equation}
is satisfied if and only if
\begin{equation}
\label{intcond}
G''(v) + \frac{6 G'(v)}{u-v} + \frac{12 G(v)}{(u-v)^2}
= G''(u) - \frac{6 G'(u)}{u-v} + \frac{12 G(u)}{(u-v)^2} \ .
\end{equation}
Integrating \eqref{partz}, we get
\begin{equation}
s z = b(v) + \frac{\el^2 G'(v)}{(u-v)} + \frac{2\el^2 G(v)}{(u-v)^2}
= c(u) - \frac{\el^2 G'(u)}{(u-v)} + \frac{2\el^2 G(u)}{(u-v)^2} \ ,
\end{equation}
where $b(\xi)$ and $c(\xi)$ are two functions.
Using the integrability condition \eqref{intcond} we see that
\begin{equation}
\begin{array}{rcl}
s( z - z_0 )
&=& \ds
\frac{\el^2 G''(v)}{6}+ \frac{\el^2 G'(v)}{(u-v)} + \frac{2\el^2 G(v)}{(u-v)^2}
\\[4mm]
&=& \ds
\frac{\el^2 G''(u)}{6} - \frac{\el^2 G'(u)}{(u-v)} + \frac{2\el^2 G(u)}{(u-v)^2} \ ,
\end{array}
\end{equation}
where $z_0$ is a constant.
Now, if we take $G(\xi)$ to be of the form
\begin{equation}
G(\xi) = a_0 + a_1 \xi + a_2 \xi^2 + a_3 \xi^3 + a_4 \xi^4  \ ,
\end{equation}
one can check that \eqref{intcond} is obeyed.

Note that $e^{2\nu}$ and $\zeta(u,v)$ in \eqref{genrzuv}
are connected through the formula
\begin{equation}
\label{nuzeta}
\frac{\zeta(u,v)}{e^{2\nu}} = \frac{\el^4}{(u-v)^4} \left[
G(u) \left( G'(v) + \frac{4 G(v)}{u-v} \right)^2
- G(v) \left( G'(u) - \frac{4 G(u)}{u-v} \right)^2 \right] \ .
\end{equation}
This formula is useful for finding $e^{2\nu}$.

Consider now the particular choice of $G(\xi)$
\begin{equation}
G(\xi) = (1-\xi^2)(1 + c \xi) \ ,
\end{equation}
which correspond to the case of the five-dimensional
black ring metric described in Section \ref{s:BRsol}.
If we take $s=1$ and $z_0=1/6$ we get
\begin{equation}
\label{findz}
z = \frac{\el^2 (1-uv)(2 + c u + c v)}{(u-v)^2} \ .
\end{equation}
We now look for constants $q$ and $\beta$ that solves the equation
\begin{equation}
r^2 + \left( z - q \el^2 \right)^2
= \frac{\el^4 \left[ \beta - c uv - 2q (u+v) \right]^2}{(u-v)^2} \ .
\end{equation}
There are precisely three solutions to this equation:
\begin{equation}
q = - c \ , \ \beta = 2 + c \ \ ; \ \ \ \ \ \
q = c \ , \ \beta = -2 + c \ \ ; \ \ \ \ \ \
q = 1 \ , \ \beta = - c \ .
\end{equation}
Write now
\begin{equation}
\label{appdefRszs}
R_i = \sqrt{r^2 + (z-z_i)^2} \spa
z_1 = - c \el^2  \spa
z_2 = c \el^2  \spa
z_3 = \el^2 \ .
\end{equation}
We get
\begin{equation}
\label{R1R2R3}
\begin{array}{c} \ds
R_1 = \frac{\el^2\left[ 2 + c ( 1 + u + v  - uv ) \right]}{(u-v)}
\spa
R_2 = \frac{\el^2\left[ 2 + c ( - 1 + u + v + uv) \right]}{(u-v)} \ ,
\\[4mm] \ds
R_3 = \frac{\el^2\left[ -c (1+ uv) - (u+v) \right]}{(u-v)} \ .
\end{array}
\end{equation}
We have furthermore that
\begin{eqnarray}
\label{Rszs}
R_1 + z - z_1 &=& \frac{2\el^2 (1+u)(1-v)(1+cu)}{(u-v)^2} \spa
R_1 - z + z_1 = \frac{2\el^2 (1-u)(-1-v)(1+cv)}{(u-v)^2} \ ,
\nn \\
R_2 + z-z_2 &=& \frac{2\el^2 (1+u)(1-v)(1+cv)}{(u-v)^2}  \spa
R_2 - z+z_2 = \frac{2\el^2 (1-u)(-1-v)(1+cu)}{(u-v)^2} \ ,
\nn \\
R_3 + z-z_3 &=& \frac{2\el^2(1-u^2)(1+cv)}{(u-v)^2} \spa
R_3 - z+z_3 = \frac{2\el^2(v^2-1)(1+cu)}{(u-v)^2} \ .
\end{eqnarray}
Finally, we can solve for $u$ and $v$ to obtain
\begin{equation}
\label{uvfromrz}
\begin{array}{l} \ds
u = \frac{(1-c)R_1 - (1+c)R_2 - 2R_3
+ 2(1-c^2)\el^2}{(1-c)R_1+(1+c)R_2+2cR_3} \ ,
\\[4mm] \ds
v = \frac{(1-c)R_1 - (1+c)R_2 - 2R_3
- 2(1-c^2)\el^2}{(1-c)R_1+(1+c)R_2+2cR_3} \ .
\end{array}
\end{equation}

%%%%%%%%%%%%%%%%%%%%%%%%%%%%%%%%%%%%%%%%%%%%%
\end{appendix}

\addcontentsline{toc}{section}{References}

%The following two lines is for bibtex only:
%\bibliographystyle{c:/mytex/INPUT/utphys}
%\bibliography{c:/mytex/BIB/bibrot,c:/mytex/BIB/biblioniels,c:/mytex/BIB/mybib}
%\bibliographystyle{../INPUT/utphys}
%\bibliography{../BIB/bibrot,../BIB/biblioniels,../BIB/mybib}

\providecommand{\href}[2]{#2}\begingroup\raggedright\endgroup

\end{document}